%% file: main.tex
\documentclass[twocolumn,showpacs,floats,floatfix,superscriptaddress,aps,prl]{revtex4-1}

\input{tex/packages}
\input{tex/definitions}

\begin{document}

\input{tex/authors}

\date{\today}

\input{tex/main/abstract}

\maketitle

\input{tex/main/introduction}


\input{tex/main/theory}

\input{tex/main/measurement}

\input{tex/main/conclusion}

\input{tex/main/acknowledgments}

\input{main.bbl}

\end{document}


\input{tex/authors}
\title{Supplemental Material to ``\basetitle'' 
}

\date{\today}

\maketitle

\input{tex/supp/supp_info}

\input{tex/supp/supp_measurement}

\bibliography{references}

%% file: tex/packages.tex
\usepackage{amsfonts,amssymb,amsmath}
\usepackage{color,calc}
\usepackage{graphicx}
\usepackage{siunitx}
\usepackage{bm}
\usepackage{hyperref}

%% file: tex/definitions.tex
\DeclareSIUnit{\gauss}{G}

\newcommand{\measurementtime}{\ensuremath{\SI{18000}{\second}}}
\newcommand{\measurementsweeps}{\ensuremath{597894}}
\newcommand{\sequencelength}{\ensuremath{\SI{15.063}{\micro \second}}}

\newcommand{\phaseacqreps}{\ensuremath{1991}}
\newcommand{\signalfrequency}{\ensuremath{\SI{1}{\mega \hertz}}}

\newcommand{\qdynefreq}{\ensuremath{\SI{4.2954 +- 0.0013}{\kilo \hertz}}}
\newcommand{\fwhm}{\ensuremath{\SI{59.9 +- 2.1}{\hertz}}}

\newcommand{\nuclearpolarization}{\ensuremath{\SI{89.1 +- 1.8}{\percent}}}

\newcommand{\densitybeforemone}{\ensuremath{0.109}}

\newcommand{\densitybeforezero}{\ensuremath{0.285}}

\newcommand{\densitybeforeone}{\ensuremath{0.606}}

\newcommand{\bzero}{\ensuremath{\SI{2043.763 +- 0.006}{\gauss}}}

\newcommand{\testsignalstrength}{\ensuremath{\SI{14.28 +- 0.04}{\micro \tesla}}}

\newcommand{\laserrepetitions}{\ensuremath{1050}}
\newcommand{\Ttilde}{\ensuremath{\SI{15.82}{\milli \second}}}

\newcommand{\sensitivity}{\ensuremath{\SI{55.64}{\micro\tesla \per \sqrt{\hertz}}}}
\newcommand{\numnvs}{\ensuremath{1000}}
\newcommand{\ensemblemeasurementsweeps}{\ensuremath{598}}
\newcommand{\ensemblemeasurementtime}{\ensuremath{\SI{18}{\second}}}
\newcommand{\ensemblesensitivity}{\ensuremath{\SI{1.76}{\micro\tesla \per \sqrt{\hertz}}}}

\newcommand{\measurementtimefactorcs}{\ensuremath{999}}

\newcommand{\mizero}{\ensuremath{m_\text{I} = 0}}

\newcommand{\mimone}{\ensuremath{m_\text{I} = -1}}

\newcommand{\cnotn}{\ensuremath{\text{C}_\text{n}\text{NOT}_\text{e}}}
\newcommand{\cnote}{\ensuremath{\text{C}_\text{e}\text{NOT}_\text{n}}}
\newcommand{\basetitle}{Quantum Memory Enhanced Multipoint Correlation Spectroscopy for Statistically Polarized NMR}

\def\i{i}

\def\i{\text{i}}

%% file: tex/authors.tex
\title{\basetitle}

\author{Tobias Spohn}
\thanks{Corresponding author: \href{mailto:kai.spohn@uni-ulm.de}{kai.spohn@uni-ulm.de}}
\affiliation{Institute of Quantum Optics and Center for Integrated Quantum Science and Technology (IQST), Ulm University, Albert-Einstein-Allee 11, 89081 Ulm, Germany}

\author{Nicolas Staudenmaier}
\affiliation{Institute of Quantum Optics and Center for Integrated Quantum Science and Technology (IQST), Ulm University, Albert-Einstein-Allee 11, 89081 Ulm, Germany}

\author{Philipp J. Vetter}
\affiliation{Institute of Quantum Optics and Center for Integrated Quantum Science and Technology (IQST), Ulm University, Albert-Einstein-Allee 11, 89081 Ulm, Germany}

\author{Timo Joas}
\affiliation{Institute of Quantum Optics and Center for Integrated Quantum Science and Technology (IQST), Ulm University, Albert-Einstein-Allee 11, 89081 Ulm, Germany}

\author{Thomas Unden}
\affiliation{NVision Imaging Technologies GmbH, Ulm D-89081, Germany}

\author{Ilai Schwartz}
\affiliation{NVision Imaging Technologies GmbH, Ulm D-89081, Germany}

\author{Philipp Neumann}
\affiliation{NVision Imaging Technologies GmbH, Ulm D-89081, Germany}

\author{Genko Genov}
\affiliation{Institute of Quantum Optics and Center for Integrated Quantum Science and Technology (IQST), Ulm University, Albert-Einstein-Allee 11, 89081 Ulm, Germany}

\author{Fedor Jelezko}
\affiliation{Institute of Quantum Optics and Center for Integrated Quantum Science and Technology (IQST), Ulm University, Albert-Einstein-Allee 11, 89081 Ulm, Germany}

%% file: tex/main/abstract.tex
\begin{abstract}

Nuclear magnetic resonance spectroscopy with solid-state spin sensors is a promising pathway for the detection of nuclear spins at the micro- and nanoscale. 
Although many nanoscale experiments rely on a single sensor spin for the detection of the signal, leveraging spin ensembles can enhance sensitivity, particularly in cases in which the signal merely originates from statistically polarized nuclear spins. 
In this work, we introduce multipoint correlation spectroscopy,
that combines the advantages of two well-established methods -- correlation spectroscopy and quantum heterodyne detection -- to enable temporally efficient measurements of statistically polarized
samples at the nanoscale with spin ensembles. We present a theoretical framework for this approach and demonstrate an experimental proof of concept with a nitrogen vacancy center in diamond. We achieve single hertz uncertainty in the estimated signal frequency, highlighting the potential applications of the technique for nanoscale nuclear magnetic resonance. 

\end{abstract}

%% file: tex/main/introduction.tex
Nuclear magnetic resonance (NMR) has been widely used for multiple applications in physics, chemistry, biology, and medicine for the past several decades. 
However, detection of a signal typically requires significant thermal polarization of a nuclear spin sample, which generally exceeds several $\si{\micro \meter ^3}$ \cite{Munuera2023prl, allert2022advances, alsina2024j}.
The thermal polarization increases when stronger bias magnetic fields are applied, rendering high magnetic fields common for NMR experiments.
Solid state spin sensors, such as the negatively charged nitrogen-vacancy (NV) center in diamond, allow for a significant reduction of the sensing volume to a few $\mathrm{nm^3}$~\cite{Mamin2013science,Staudacher2013science, schwartz2019blueprint}. Then, statistical polarization becomes dominant, especially at low magnetic fields \cite{Staudenmaier2022, Staudenmaier2023prl}.
In this regime, the polarization follows Poissonian random statistics $\propto N_\mathrm{nuc}^{-1/2}$ and is dependent only on the number of nuclear spins $N_\mathrm{nuc}$ in the sensing volume and no longer on the bias magnetic field \cite{Mamin2013science,Staudacher2013science}.
This enables applications such as single molecule  \cite{du2024single} or spin \cite{sushkov2014magnetic} spectroscopy and the determination of nanoscale structure at ambient conditions \cite{wu2016diamond, devience2015nanoscale}.

\begin{figure}[t!]
    \centering
    \includegraphics[width=\linewidth]{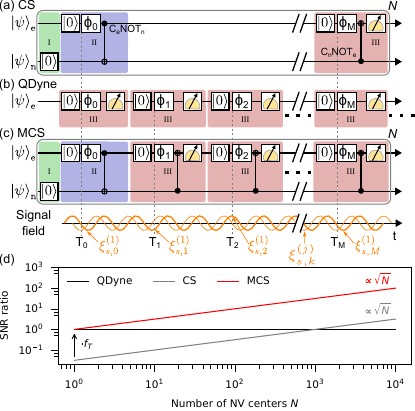}
    \caption{(a) CS measurement scheme with ancilla based memory.
    (I) The memory spin $n$ is initialized before (II) acquiring an initial phase $\phi_0^{(j)}$ with the sensor spin $e$, where $(j)$ denotes the $j$-th repetition of the experiment or the $j$-th NV in an ensemble, $j=1,\dots, N$  ($j$ omitted for simplicity in the figure) and mapping it onto the memory spin by using a $\cnote$ gate. 
    In a subsequent phase measurement (III) the phase $\phi^{(j)}_k$ $\left( k \in \{ 1,...,M\} \right)$ is acquired, which is correlated to the initial phase by performing a $\cnotn$ gate. The phases $\phi_k^{(j)}$ depend on the signal phase $\xi^{(j)}_{\mathrm{s}, k}$ at the beginning of the respective phase acquisition. 
    (b) QDyne measurement scheme.
    Only a sensor spin $e$ is used to acquire the phases $\phi_k$ at a constant rate.
    Commonly, the protocol is performed for more than $M+1$ phase acquisitions in order to improve the SNR.
    (c) MCS scheme.
    Similar to CS, a sensor and memory spin are used to establish correlations between an initial phase $\phi_0$ and the measurement phases $\phi_k$ by storing and retrieving $\phi_0$ on the memory spin.
    $M+1$ phase acquisitions are performed during one sequence, which can be repeated $N$ times.
    (d) Comparison of the SNR of the three protocols vs. the number $N$ of NV centers in an ensemble, where photon shot noise is assumed to be the main noise source. 
    SNR of MCS and CS (QDyne) protocols are normalized to the one of MCS (QDyne) with a single NV center. 
    $\text{SNR}_{\text{Qdyne}}$ does not change with $N$ while SNR of both CS and MCS $\propto \sqrt{N}$.
    MCS is favorable by factor $f_T\propto \sqrt{M}$ to CS as its data acquisition lacks long waiting times.}
    \label{fig:protocols}
    \end{figure}

Several measurement protocols have been established to detect nano- and microscale NMR signals with NV centers, most prominently correlation spectroscopy (CS) \cite{Laraoui2013natcommun, Staudacher2015natcommun, Kong2015prapplied,ZaiserNatComm2016, PfenderNatComm2017}, quantum heterodyne detection (QDyne) \cite{SchmittScience2017, Boss2017science, Staudenmaier2023prl}, and coherently averaged synchronized readout (CASR) \cite{Glenn2018CASR, Arunkumar2021PHIPCASR, Neuling2023}.
CS correlates the phases $\phi_0, \phi_k$ acquired at times $T_0, T_k$, respectively, by a quantum sensor during two sensing periods, as shown in \autoref{fig:protocols}~(a).
The phase $\phi_0$ is stored in the populations of the quantum sensor \cite{Laraoui2013natcommun, Staudacher2015natcommun, Kong2015prapplied} or an ancillary memory qubit \cite{ZaiserNatComm2016, PfenderNatComm2017,Rosskopf2017}.
By sweeping $k$, the autocorrelation of the accumulated phases $\phi_0$ and $\phi_k$ is detected.
The frequency resolution of this method is limited by the spin lifetime of the NV center \cite{Laraoui2013natcommun, Staudacher2015natcommun, Kong2015prapplied}, or that of the ancillary memory qubit \cite{ZaiserNatComm2016, PfenderNatComm2017,Rosskopf2017}, where the latter can be several orders of magnitude longer \cite{ZaiserNatComm2016}, leading to improved resolution.
A major disadvantage is the typically long idle time between the two phase acquisition periods when no data are acquired. 
Furthermore, multiple measurements  have to be performed, where the idle waiting time is varied, in order to obtain the phases correlation function. 

QDyne mitigates this problem, by repeating the measurement at a constant rate, determined by the time needed for initialization, detection, and readout \cite{SchmittScience2017, Boss2017science}, as visualized in \autoref{fig:protocols}~(b). 
This significantly reduces the temporal overhead of the measurement, resulting in efficient signal acquisition.
The correlations are established indirectly 
and are retrieved in post-processing. 
However, Qdyne is limited by the low readout efficiency for single NV centers due to photon shot noise \cite{Staudenmaier2023prl}. 
The experiments then require a longer measurement time to reach a sufficient signal-to-noise ratio (SNR), limiting sensitivity. 

Using multiple NV centers in an ensemble simultaneously can significantly improve sensitivity \cite{barry2020sensitivity}. 
However, the SNR of QDyne in nanoscale NMR does not improve with a larger number of NV centers. 
While the photon shot noise decreases, 
the detected signal also drops with statistical polarization.
Specifically, the photoluminescence signal from the $j$-th NV center is linearly dependent on its acquired phase $\phi_k^{(j)}$, which varies for each NV center as they have different sensing volumes.
As all NV centers are read out simultaneously, averaging is performed over all these phases.
This leads to an effective increase in the sensing volume, resulting in a reduction in the signal detected from statistical polarization \cite{Mamin2013science,Staudacher2013science}. 
As a result, there is typically no net benefit when using QDyne with an NV ensemble \cite{Supplemental}. In contrast, CS, with its direct detection of phase correlations, 
detects a signal, which is proportional to the average phase variance for each NV center. Thus, the effective sensing volume does not increase, making CS applicable to NV ensembles \cite{Staudenmaier2022}.

In this Letter, we introduce multipoint correlation spectroscopy (MCS), which combines the advantages of QDyne and CS, i.e. the high data acquisition rate of the former and CS's improvement of sensitivity with multiple NV centers without the loss of statistical polarization signal.
Specifically, MCS detects the correlation of an initial phase $\phi_0$ with many subsequent phases $\phi_k$, acquired in a QDyne-type fashion. 
By storing the information about $\phi_0$ in a memory qubit, the following phases $\phi_k$ can be correlated with $\phi_0$. The protocol is limited by the lifetime of the memory qubit, which can reach up to a second for the intrinsic nitrogen nuclear spin \cite{Neumann2010singleshot}, allowing for single-hertz resolution.

%% file: tex/main/theory.tex
\emph{Theory.}--- 
The theoretical description of MCS is general, as it is applicable to any pair of a sensor $(e)$ and a memory $(n)$ qubit for advanced quantum sensing. \autoref{fig:protocols} shows a comparison of CS, QDyne and MCS. 
In our particular experiment, we use a negatively charged NV center in diamond with its electron spin  -- the sensor qubit, and the nuclear spin of nitrogen -- the memory qubit. 
To match its description, the states are labeled as $\text{$|m_\text{s}$}\rangle _e\text{$|m_\text{I}$}\rangle_n$ corresponding to $\{\text{$|$0}\rangle _e\text{$|$0}\rangle_n,\text{$|$0}\rangle _e\text{$|$+1}\rangle_n, \text{$|$-1}\rangle _e\text{$|$0}\rangle_n, 
 \text{$|$-1}\rangle_e\text{$|$+1}\rangle_n\}$, 
(see \autoref{fig:levelscheme} and \cite{Supplemental}). 
The system is initialized in state $\text{$|$0}\rangle _e\text{$|$0}\rangle_n$ (step I in \autoref{fig:protocols} (c)).
For simplicity of presentation, initialization is assumed as ideal. 
Imperfect initialization would lead to reduced contrast \cite{Supplemental}. 

MCS then includes an initial phase acquisition (step II), where a signal-dependent phase $\phi_0$ is accumulated and subsequently stored as a population difference in the $^{14}$N nuclear spin memory qubit. 
We apply a $\pi/2$ pulse on the sensor spin to create a coherent superposition state.
Then we use dynamical decoupling (DD) with the XY8 \cite{Gullion1990jmr} sequence to increase the coherence time and allow for the detection of weak oscillating signals \cite{DegenRMP2017}. Robust DD sequences \cite{Gullion1990jmr,Suter2016RevModPhys,GenovPRL2017,GenovPRR2020,Louzon2025prl} allow phase acquisition with NV ensembles where frequency and amplitude inhomogeneities might be present \cite{Supplemental}. 
We apply DD on resonance, i.e., the time between the centers of the $\pi$ pulses is $\tau\approx 1/(2\nu_s)$ \cite{DegenRMP2017}, where $\nu_s=\omega_s/(2\pi)$ is the frequency of the signal 
and $\omega_s$ its angular frequency. 
In case of nano-NMR, based on statistical polarization, $\nu_s$ is the expected Larmor frequency of the nuclear spins ensemble \cite{Staudenmaier2022}. 
The accumulated phase is \cite{DegenRMP2017}
\begin{equation}
\phi_0^{(j)}\approx\frac{2}{\pi}g_{s,0}^{(j)}t_{\text{DD}}\cos{(\xi_{s,0}^{(j)})}=\frac{2}{\pi}\gamma_{\text{NV}}B_{s,0}^{(j)}t_{\text{DD}}\cos{(\xi_{s,0}^{(j)})},
\end{equation}
where $g_{s,0}^{(j)}$ is the signal amplitude, which is proportional to the signal magnetic field envelope $B_{s,0}^{(j)}$, $t_{\text{DD}}$ is the XY8 duration, and $\xi_{s,0}^{(j)}$ is the signal phase at time $T_0$ when detection starts. 
The superscript $j \in \{1,...,N\}$ denotes the respective value for the $j$-th sensor in an ensemble or the $j$-th experimental run when the measurement is performed on a single spin pair. 
The phase information $\phi_0^{(j)}$ is then transferred to the populations of the sensor spin states by a $\pi/2$ Y pulse, 
and is subsequently stored in the population of the memory qubit by a $\pi$ pulse on the memory qubit, conditioned on the spin state of the sensor, which we label a $\cnote$ gate.  

In step III, we perform a series of subsequent phase measurements.
In each experiment, the sensor is prepared in the state $|0\rangle_e$, ideally without affecting the memory qubit. 
Then, we repeat the previous step and the sensor accumulates a phase
\begin{equation}
\phi_k^{(j)}\approx\frac{2}{\pi}g_{s,k}^{(j)}t_{\text{DD}}\cos{(\xi_{s,k}^{(j)})}
\end{equation}
where $g_{s,k}^{(j)}=\gamma_{\text{NV}}B_{s,k}^{(j)}$ is the signal amplitude during the $k$-th period.  
The signal has the phase $\xi_{s,k}^{(j)} = \xi_{s,0}^{(j)} + \omega_s \Delta T_k$ at the beginning of $k$-th detection with $\Delta T_k=T_k-T_0$. 
Information about the initial period phase $\phi_0^{(j)}$ is retrieved after every phase acquisition with a $\pi$ pulse on the sensor spin, conditioned on the memory spin state, which we label $\cnotn$ gate.
The population of the sensor spin state $|0\rangle_e$ and the detected signal depends on the correlation between $\phi_0^{(j)}$ and $\phi_k^{(j)}$ \cite{Supplemental} 
\begin{align}\label{eq:signal}
S_k &\approx \eta\left(1+\frac{c}{2}\langle\phi_0^{(j)}\phi_k^{(j)}\rangle\right)\notag\\
&=\eta\left(1+\frac{c}{2}\frac{4}{\pi^2}\gamma_{\text{NV}}^2 B_{\text{rms}}^2t_{\text{DD}}^2\cos{\left(\omega_s \Delta T_k \right)}\right),
\end{align}
where $\eta=(\eta_0+\eta_1)/2$ is the average signal, typically number of photons, with $\eta_0$ ($\eta_1$) the expected signal when the sensor qubit is in state $|0\rangle$ ($|\text{-1}\rangle$), while $c=(\eta_0-\eta_1)/\eta$ is the relative experimental contrast. The phase $\phi_k^{(j)}$ is assumed to be small, so that $\sin{(\phi_k^{(j)})}\approx \phi_k^{(j)}$ for the first equality. 
The averaging $\langle \cdot \rangle $ is done on the $j=1,\dots, N$ different sensors
in the ensemble, or over the $N$ repetitions of the experiment on a single spin pair. 
The quantity $B_{\text{rms}}^2=B_s^2/2$ for a classical signal with a constant amplitude $B_{s,0}^{(j)}=B_{s,k}^{(j)}=B_s$ for all measurements (or all spin pairs in an ensemble), or $B_{\text{rms}}$ is the root mean square of the magnetic field due to statistical polarization, sensed by a sensor spin in an ensemble \cite{Staudenmaier2022,Staudenmaier2023prl,Supplemental}. 
The signal obtained after every $k$-th measurement allows extraction of the correlation between the accumulated phase during the $k$-th measurement and during the initial measurement. The signal scales as $S_k\sim \eta \sim N$, while the usually dominant photon shot noise $\sim \sqrt{N}$, leading to improved $\text{SNR}\sim \sqrt{N}$, as shown in Fig. \ref{fig:protocols} (d).  
This is a significant advantage over Qdyne with spin ensembles, where the signal from the $k$-th measurement depends on $\langle\phi_k^{(j)}\rangle\sim B_{\text{rms}}/\sqrt{N}$ for $ j=1,\dots, N$, meaning that $\text{SNR}_{\text{Qdyne}}$ does not improve for ensembles with $N$ spins compared to the single spin case \cite{Supplemental}.
Although the signal oscillates with an angular frequency $\omega_s$, measurements at times $T_k$ typically detect an undersampled frequency $\omega_\mathrm{u} = (2 \pi) f_\mathrm{u}$, related to $\omega_s$ \cite{SchmittScience2017,Staudenmaier2022,Staudenmaier2023prl}, which can be extracted by performing a Fourier transform. As the detected signal is related to the autocorrelation function of the sensed magnetic field, the Fourier transform produces its power spectral density \cite{Supplemental}. 
The line width is 
$\sim 1/T_\text{total}\approx 1/\Delta T_M\,,$
where $M$ is the total number of phase acquisitions after the initial phase accumulation. 
The total measurement time $T_\text{total}$ is limited by the lifetime $T_{1,\mathrm{nuc}}$ of the memory spin qubit, considering that it could be affected by the initialization of the sensor spin \cite{Supplemental}. 
The key advantage of MCS over CS is an improved SNR by factor
\begin{equation}
f_T=\sqrt{\frac{M}{2}\left(1+\frac{1+T/T_{\text{init}}}{1+M T/T_{\text{init}}}\right)}\,,    
\end{equation}
where $T_{\text{init}}$ is the total duration of steps I and II and $T=T_{k+1}-T_{k}$ is the duration of one measurement in step III \cite{Supplemental}. 
This improved SNR is caused by reducing measurement idle time between the two phase acquisition periods of CS and replacing it with repeated phase measurements.

%% file: tex/main/measurement.tex
\begin{figure}[t!]
    \centering
    \includegraphics[width=\linewidth]{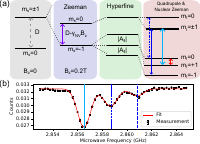}
    \caption{(a) Level scheme of an NV center.
    Electronic spin levels $m_\mathrm{s} = 0, \pm 1$ are split by the zero field splitting $D$ and Zeeman interaction $\pm \gamma_\mathrm{NV} B_z$, creating non-degenerate spin levels. 
    Only the subsystem $m_\mathrm{s} = 0, -1$ is considered.
    Hyperfine $A$, quadrupolar $P$ and nuclear Zeeman $\gamma_\mathrm{N} B_z$ interaction with the intrinsic $^{14}\mathrm{N}$ nuclear spin splits the electron spin states into three additional sublevels $m_\mathrm{I} = 0, \pm 1$.
    (b) Pulsed ODMR measurement reveals the hyperfine structure of the NV center with intrinsic \textsuperscript{14}N for the $m_\mathrm{s} = 0 \leftrightarrow m_\mathrm{s} = -1$ transition.
    $\nuclearpolarization$ of the nuclear spin population is in the $m_\mathrm{I}=0, +1$ states. 
    The dashed lines show the corresponding microwave transitions in (a), which are separated by the hyperfine coupling of approximately 2.1 MHz.
    The protocol was demonstrated, using the red (light blue) transition for RF (selective and strong MW) pulses.}
    \label{fig:levelscheme}
\end{figure}

\begin{figure*}[t!]
    \centering
    \includegraphics[width=\linewidth]{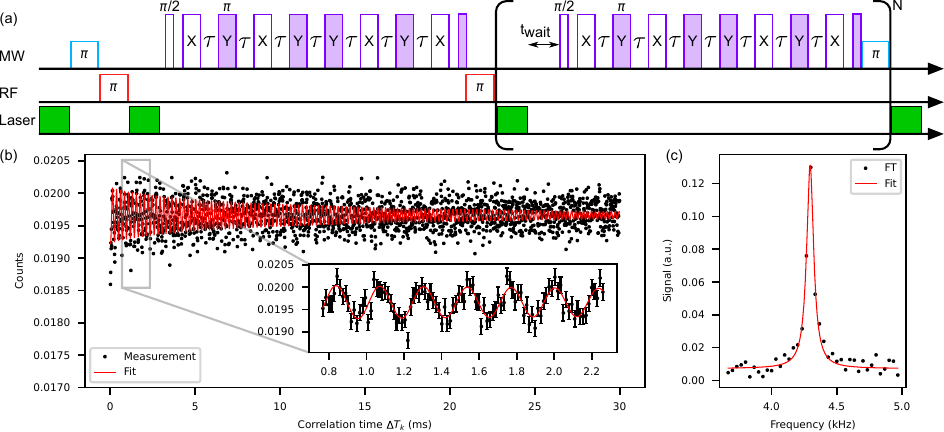}
    \caption{(a) Experimental scheme of MCS with an NV center in diamond. 
    First, the nitrogen nuclear spin is initialized using selective MW (blue) and RF (red) $\pi$ pulses. 
    The signal is then probed for the first time, using XY8-1 (purple). 
    The acquired phase is mapped onto the nitrogen nuclear spin, by utilizing an RF $\pi$-pulse.
    Subsequently, the phase acquisition is performed $M = \phaseacqreps$ times, correlating the acquired phase to the phase stored in the population of the nitrogen nuclear spin by applying a selective MW pulse.
    The sequence is repeated for $N=\measurementsweeps$ times.
    (b) Experimentally measured NV center fluorescence response for a $\testsignalstrength$ signal at $\SI{1}{\mega \hertz}$ when running the sequence in (a). 
    Error bars have been omitted for visiblity purposes and the inset shows a zoom in on the data, with error bars corresponding to photon shot noise.
    (c) Power spectral density of the measured signal.
    A fit to the signal reveals the detected undersampled frequency $f_\mathrm{u} = \qdynefreq$ with a full width at half maximum of $\fwhm$ corresponding to a resolution limit due to the nuclear spin lifetime $\widetilde{T}_{1, \mathrm{nuc}}$ under laser illumination.}
    \label{fig:results}
\end{figure*}

\emph{Experimental Demonstration.}--- 
We demonstrate the applicability of the protocol with a single NV center with an intrinsic \textsuperscript{14}N nuclear spin. 
Its level structure at a magnetic field of $\bzero$ is shown in \autoref{fig:levelscheme} (a) with only the subsystem $m_\text{s}=0, -1$ being used as the sensing qubit. 
Measurements are carried out using a home-built confocal microscope, using the Qudi measurement suite \cite{Binder2017qudi}. 
The MCS pulse sequence for an NV center is shown in \autoref{fig:results} (a).
First, we perform optical pumping with a green laser pulse to initialize the electron spin of the NV center in $\text{$|$0}\rangle _e$. 

Using weak MW pulses, the hyperfine structure of the nitrogen nuclear spin becomes visible in an ODMR spectrum, as shown in \autoref{fig:levelscheme} (b).
This spectrum can be used to determine the degree of polarization of the nitrogen nuclear spin by comparing the area under the three Gaussians the data is fit with.
The populations of nuclear spin states after long laser irradiation are thus
$P_{\text{$|$0}\rangle _e\text{$|$+1}\rangle_n}=\densitybeforeone$, $P_{\text{$|$0}\rangle _e\text{$|$0}\rangle_n}=\densitybeforezero$, and $P_{\text{$|$0}\rangle _e\text{$|$-1}\rangle_n}=\densitybeforemone$ with the corresponding drops in fluorescence denoted from left to right \cite{Supplemental}. 
The goal is to prepare the system in $\text{$|$0}\rangle _e\text{$|$0}\rangle_n$. 
A selective first MW (subsequent RF) $\pi$ pulse is applied on the transition $\text{$|$0}\rangle _e\text{$|$+1}\rangle_n \leftrightarrow \text{$|$-1}\rangle _e\text{$|$+1}\rangle_n$ ($\text{$|$-1}\rangle _e\text{$|$+1}\rangle_n \leftrightarrow \text{$|$-1}\rangle _e\text{$|$0}\rangle_n$). Then we use optical pumping to reinitialize the electron spin state. 
In the end, the population of the $m_I=+1$ state (left dip of \autoref{fig:levelscheme} (b)) is ideally pumped to $m_I=0$ (central dip) \cite{Supplemental}. 
The procedure can be modified to pump the population of the $m_I=-1$ state to $m_I=0$, which is not done in this experiment, due to the low population of the former.  

\autoref{fig:results} (b) shows the fluorescence response of the NV center for the MCS protocol, using a test signal of frequency $\signalfrequency$ and amplitude $\testsignalstrength$. 
The measured signal oscillates with the undersampled frequency $f_\mathrm{u} = \qdynefreq$, which we estimate with a precision reaching the single hertz level.
The resulting full width at half maximum of the power spectral density, shown in \autoref{fig:results} (c), is $\fwhm$.
This frequency resolution is directly limited by the nuclear spin lifetime $\widetilde{T}_{1, \mathrm{nuc}}$ of the intrinsic nitrogen nuclear spin.
During laser irradiation, the intrinsic nuclear spin of the NV center gradually depolarizes \cite{Neumann2010singleshot}, limiting its lifetime.
A nuclear polarization decay time of $\widetilde{T}_{1, \mathrm{nuc}} \approx M_{\mathrm{limit}} T = \Ttilde$ is detected during the measurement, where $T = T_{k+1} - T_k = \sequencelength$ is the time between two subsequent phase measurements and $M_\mathrm{limit} = \laserrepetitions$ is the number of lasers, leading to a $1/e$ decay of the nuclear spin polarization \cite{Supplemental}.
A line width of $\sim 1/\Delta T_M\approx \SI{33}{\hertz}$ would be expected without the effect of laser initialization on the memory spin.
In the measurement, the number of acquired phases is $M=\phaseacqreps$, confirming that the protocol is limited by the shortened nuclear spin lifetime. 
In order to limit the effect of this decay, a waiting time $t_\mathrm{wait}$ may be introduced between the $M$ phase acquisitions,
reducing the number of laser pulses per unit time and thus prolonging the nuclear $\widetilde{T}_{1, \mathrm{nuc}}$ time, which results in a better frequency resolution.
This allows to increase the maximum phase acquisition time $\Delta T_M$ of the protocol, resulting in a higher frequency resolution.
It can also be improved by increasing the static bias magnetic field $B_0$, as
the lifetime of the nitrogen nuclear spin under laser illumination depends quadratically on the bias magnetic field \cite{Neumann2010singleshot, Supplemental}.

The measurement is acquired for $\measurementtime$ or $N=\measurementsweeps$ repetitions, resulting in a sensitivity of $\sensitivity$.
As described in the theory section, this sensitivity can be directly improved by using an NV ensemble to acquire already averaged data within one measurement run.
Assuming that all parameters remain constant, using, e.g. $\numnvs$ NV centers instead of one, the total measurement time is shortened to $\sim \ensemblemeasurementtime$ or $N \approx \ensemblemeasurementsweeps$ repetitions, improving the sensitivity to $\ensemblesensitivity$.

Sensitivity can also be optimized by improving the readout signal and contrast, as shown in \autoref{eq:signal} \cite{barry2020sensitivity}.
Another limitation is the imperfect initialization of the NV center, which leads to reduced nuclear spin polarization and a limited readout contrast~\cite{Supplemental}.
The complete polarization of the electron \cite{wirtitsch2023exploiting, hopper2020real} and nuclear memory spin would thus enhance sensitivity.
A second initialization step, transferring the $\mimone$ population to $\mizero$ could also improve nuclear spin polarization.
Using an NV center with a \textsuperscript{15}N nucleus ($I=1/2$) would eliminate the need for a second initialization step.

The improved SNR of MCS compared to CS becomes evident when calculating the time required to acquire the data in \autoref{fig:results} (b) using CS.
A CS measurement would take $\sim \measurementtimefactorcs$ times longer \cite{Supplemental}, due to the large measurement idle times and the inability to record the $N$ data points during a single sequence run.
This approximately three orders of magnitude improvement in measurement time underscores the temporal advantage MCS has over CS, as the SNR of MCS would be improved by a factor of $f_T \approx \sqrt{\measurementtimefactorcs}$ compared to CS as shown in \autoref{fig:protocols} (d).

%% file: tex/main/conclusion.tex
\emph{Conclusion.}---
We proposed theoretically and demonstrated experimentally multipoint correlation spectroscopy, which uses the intrinsic nitrogen nuclear spin as a memory qubit and combines the benefits of correlation spectroscopy and QDyne protocols for single- and spin-ensemble measurements of NMR signals. 
Using QDyne's high data acquisition rate, we significantly reduce the measurement idle time in comparison to correlation spectroscopy. 
The protocol directly probes the correlation between the accumulated phases in different detection periods, which makes it applicable to both single and ensemble NV center experiments and for detecting signals, based on thermal and statistical polarization. 
We achieve single hertz uncertainty in the estimated signal frequency in a proof-of-concept experiment in a single NV center, highlighting the broad applications of the technique. 
Liquid nano-NMR applications are limited by the diffusion of the sample molecules \cite{Staudenmaier2022, Staudenmaier2023prl}.
Our protocol does not mitigate this effect.
However, several other techniques have been demonstrated that combat diffusion by confining the NMR sample \cite{cohen2020confined, liu2022using, zheng2024observation} or optimizing the NV centers depth, taking into account the trade off between diffusion broadening and signal strength \cite{pham2016nmr}.
Combining these techniques with MCS would allow for significant improvement in the sensitivity of nano-NMR, compared to standard QDyne and correlation spectroscopy.

During the final stages of the preparation of this manuscript, we became aware of a related independent work \cite{maier2025}.

%% file: tex/main/acknowledgments.tex
\emph{Acknowledgment.}---
The authors thank Santiago Oviedo-Casado and Liam P. McGuinness for useful discussions. This work was supported by the German Federal Ministry of Research (BMBF) by future cluster QSENS and projects DE-Brill (No. 13N16207), SPINNING, DIAQNOS (No. 13N16463), quNV2.0 (No. 13N16707), QR.X and Quamapolis (No. 13N15375), DLR via project QUASIMODO (No.50WM2170), Deutsche Forschungsgemeinschaft (DFG) via Projects No. 386028944, No. 445243414, and No. 387073854 and Excellence Cluster POLiS, BMBF via project EXTRASENS (No. 13N16935), Europe   an Union’s HORIZON Europe program via projects QuantERA II (No. 101017733), QuMicro (No. 101046911), SPINUS (No. 101135699), CQuENS (No. 101135359), QCIRCLE (No. 101059999), and FLORIN (No. 101086142), European Research
Council (ERC) via Synergy grant HyperQ (No. 856432), Carl-Zeiss-Stiftung via the Center of Integrated Quantum Science and technology (IQST) and project Utrasens-Vir, as well as the Baden-W{\"u}rttemberg Foundation.

\emph{Data availability}
Some of the data that support the findings of this article are openly available \cite{plotdata}. Datasets are not publicly available. The data are available from the authors upon reasonable request.

%% file: main.bbl
%

%% file: tex/supp/supp_info.tex
\section{Theory details}\label{Appendix:Theory}

\subsection{The System}\label{SM_Subsection:System}

In order to characterize the time evolution of our system during the multipoint correlation spectroscopy (MCS) protocol, we consider the Hamiltonian of a single negatively charged NV center with an inherent nitrogen $^{14}$N nucleus in its orbital ground state $^3$A$_2$ ($\hbar=1$) \cite{Suter2017nv}.
\begin{align}\label{eq:SM_Hamiltonian_total}
H_{\text{$^{14}$NV-}}&=H_0+H_{\text{Z}}+H_{\text{hfN}}+H_{\text{Q}}+H_{\text{NZ}},\notag\\
&=\underbrace{D S_z^2}_{H_0}+\underbrace{\gamma_{\text{NV}}B_z S_z}_{H_{\text{Z}}}
+\underbrace{A_{||}S_z I_z+A_{\perp}\left(S_{+}I_{-}+S_{-}I_{+}\right)}_{H_{\text{hfN}}}
+\underbrace{P\left(I_z^2-\frac{2}{3}\mathbf{1}\right)}_{H_{\text{Q}}}-\underbrace{\gamma_{\text{N}}B_z I_z}_{H_{\text{NZ}}}\notag\\
&\approx\underbrace{D S_z^2}_{H_0}+\underbrace{\gamma_{\text{NV}}B_z S_z}_{H_{\text{Z}}}
+\underbrace{A_{||}S_z I_z}_{H_{\text{hfN}}}
+\underbrace{P\left(I_z^2-\frac{2}{3}\mathbf{1}\right)}_{H_{\text{Q}}}-\underbrace{\gamma_{\text{N}}B_z I_z}_{H_{\text{NZ}}},
\end{align}
where $H_0$ is the zero-field Hamiltonian and $D=(2\pi)2.87\,$ GHz is the zero-field splitting of the ground state of the NV$^{-}$ center electron spin. The respective spin operators $S_m$ ($I_m$), $m=x,y,z$ are defined to act on the electron (nuclear) spin states of the NV center only, e.g., $S_z$ is equivalent to $S_z\otimes \mathbf{1}$, where $\mathbf{1}$ is the identity operator for the nuclear spin. Both the electron and nuclear spin states are triplets $S=I=1$ since we consider the $^{14}N$ isotope. $H_{\text{Z}}$ is the Hamiltonian due to the Zeeman splitting of the NV center electron spin, where the magnetic field is aligned along the NV center quantization axis and has a magnitude $B_z = \bzero$, $\gamma_{\text{NV}}=(2\pi)2.803 \,$ MHz/G; $H_{\text{hfN}}$ is the Hamiltonian due to the hyperfine interaction between the NV center electron spin and the $^{14}$N nuclear spin with $A_{||}=-(2\pi)2.166$ MHz and we neglected the effect of the transverse component $A_{\perp}$ as it affects the system only to second order and is much smaller than the zero-field splitting and the Zeeman splitting at the high magnetic field we apply; $H_{\text{Q}}$ is the Hamiltonian due to the quadrupole interaction with the $^{14}$N nuclear spin $I=1$ and $P=-(2\pi)4.945$ MHz. Finally, $H_{\text{NZ}}$ is the Hamiltonian due to the Zeeman splitting of the $^{14}$N nuclear spin, with $\gamma_{\text{N}}=(2\pi)0.308 \,$ kHz/G being the gyromagnetic ratio of $^{14}$N. 

\begin{table}[t]
\caption{State labels and energy levels of the corresponding states, as calculated from the Hamiltonian in Eq. \eqref{eq:SM_Hamiltonian_total}. As an example, the label $\text{$|$+1}\rangle _e\text{$|$+1}\rangle_n$ corresponds to the $m_s=+1$, $m_I=+1$ state, as shown in Fig. \ref{fig:SM_scheme}(a).
}
\begin{tabular}{l l l} 
\hline 
State Label~~~~~~~~~~~~~~~~~~~~~~~& Energy level~~~~~~~~~~~~~~~~~~~~~~~~~~~~~~~~~~~~~~~~& Energy level $/(2\pi)$ MHz \\ 
\hline 
$\text{$|$+1}\rangle _e\text{$|$+1}\rangle_n$ & $D+\gamma_{\text{NV}}B_z +A_{||}+\frac{P}{3}-\gamma_\text{N} B_z$ & 8594.2 \\
[0.5ex]
\hline 
$\text{$|$+1}\rangle _e\text{$|$0}\rangle_n$ & $D+\gamma_{\text{NV}}B_z -\frac{2P}{3}$ & 8602 \\
[0.5ex]
\hline 
$\text{$|$+1}\rangle _e\text{$|$-1}\rangle_n$ & $D+\gamma_{\text{NV}}B_z -A_{||}+\frac{P}{3}+\gamma_\text{N} B_z$ & 8599.8 \\
[0.5ex]
\hline 
$\text{$|$0}\rangle _e\text{$|$+1}\rangle_n$ & $\frac{P}{3}-\gamma_\text{N} B_z$ & -2.28 \\
[0.5ex]
\hline 
$\text{$|$0}\rangle _e\text{$|$0}\rangle_n$ & $-\frac{2P}{3}$ & 3.30 \\
[0.5ex]
\hline 
$\text{$|$0}\rangle _e\text{$|$-1}\rangle_n$ & $\frac{P}{3}+\gamma_\text{N} B_z$ & -1.02 \\
[0.5ex]
\hline 
$\text{$|$-1}\rangle _e\text{$|$+1}\rangle_n$ & $D-\gamma_{\text{NV}}B_z -A_{||}+\frac{P}{3}-\gamma_\text{N} B_z$ & -2858.77 \\
[0.5ex]
\hline 
$\text{$|$-1}\rangle _e\text{$|$0}\rangle_n$ & $D-\gamma_{\text{NV}}B_z -\frac{2P}{3}$ & -2855.36 \\
[0.5ex]
\hline 
$\text{$|$-1}\rangle _e\text{$|$-1}\rangle_n$ & $D-\gamma_{\text{NV}}B_z +A_{||}+\frac{P}{3}+\gamma_\text{N} B_z$ & -2861.84 \\
[0.5ex]
\hline 
\end{tabular}
\label{Table:SM_energy_levels} 
\end{table}

\begin{figure*}[t!]
    \centering
    \includegraphics[width=\textwidth]{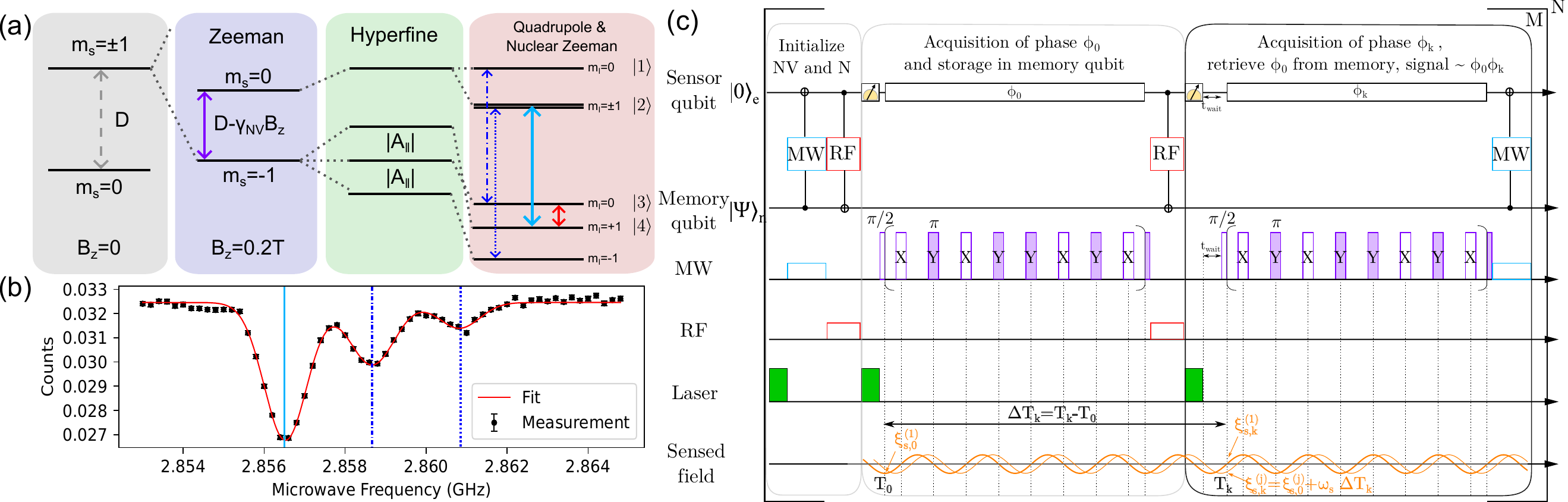}
    \caption{(a) Ground state $\left(^3\mathrm{A}_2\right)$ level scheme of an NV center.
    Electronic spin levels $m_\mathrm{s} = 0, \pm 1$ are  split by the zero field splitting $D$.
    Zeeman Interaction additionally shifts these levels by $\pm \gamma_\mathrm{NV} B_z$, detuning $m_\mathrm{s} = + 1$ far apart, thus only the subsystem $m_\mathrm{s} = 0, -1$ is considered.
    Hyperfine interaction with the intrinsic $^{14}\mathrm{N}$ nuclear spin splits $m_\mathrm{s} = -1 $ into three additional sublevels $m_\mathrm{I} = 0, \pm 1$, separated by $|A_{||}|$.
    Quadrupolar interaction $P$ and nuclear Zeeman $\pm \gamma_\mathrm{N} B_z$ shifts these hyperfine levels further.
    (b) Slow pulsed ODMR measurement to reveal the hyperfine structure of 14N.
    Fitting \autoref{eq:triple_gaussian} to the data, the degree of polarization and magnetic field strength $B_0$ can be determined.
    $\nuclearpolarization$ of the nuclear spins population is in the $m_I=-1$ and $m_I=0$ state without any polarization protocol applied. 
    Therefore, initialization is only performed between these two states.
    The magnetic field is $\bzero$. The dashed lines show the corresponding MW transitions in (a), which are separated by approximately 2.16 MHz. 
    The measurement was acquired for $\sim \SI{20}{\minute}$ with $4189424$ measurement sweep repetitions.
    (c) MCS pulse sequence.
    First, the nitrogen nuclear spin is initialized. We then acquire an initial phase $\phi_0^{(j)}\sim \cos{(\xi_0^{(j)})}$, using a XY8-1 sequence, which depends on the signal initial phase at the beginning of the phase acquisition time $T_0$. 
    The subscript $(j)$ denotes the $j$-th repetition of the experiment or the $j$-th NV center in a spin ensemble, $j=1,\dots,N$. We use $\phi_k$ in the figure for simplicity of presentation. 
    The acquired phase $\phi^{(j)}_0$ is mapped onto the populations of the nitrogen nuclear spin, by utilizing an RF $\pi$-pulse, which performs a $\cnotn$ gate.
    Subsequently, the phase acquisition is repeated $M$ times to acquire phases $\phi^{(j)}_k$, starting at times $T_k$, and each acquired phase is correlated to the phase $\phi^{(j)}_0$, stored in the population of the nitrogen nuclear spin by applying a selective MW pulse, which performs an effective $\cnote$ gate.
    The phase $\phi^{(j)}_k$ is directly dependent on the phase of the signal $\xi^{(j)}_{s, k}$ at the beginning of each phase acquisition.
    The experiment is performed $N$ times or with an NV ensemble with $N$ spins. 
    }
    \label{fig:SM_scheme}
\end{figure*}

The resulting Hamiltonian in the two-qubit basis is diagonal and the energy levels of the respective states are described in \autoref{Table:SM_energy_levels}, with the important states depicted in \autoref{fig:SM_scheme} (a). 
\autoref{fig:SM_scheme} (b) shows an optically detected magnetic resonance (ODMR) spectrum where the three dashed lines correspond to the transitions between states $\text{$|$0}\rangle _e\text{$|m_\text{I}$}\rangle_n \leftrightarrow \text{$|$-1}\rangle _e\text{$|m_\text{I}$}\rangle_n$.
The transition frequencies are given by $\gamma_{\text{NV}}B_z-D+ m_\text{I} A_{||}$ with $m_\text{I}=+1, 0, -1$ for the respective transition (from left to right). 
Specifically, the frequencies of the transitions $\text{$|$0}\rangle _e\text{$|$+1}\rangle_n \leftrightarrow \text{$|$-1}\rangle _e\text{$|$+1}\rangle_n$ (light blue), $\text{$|$0}\rangle _e\text{$|$0}\rangle_n \leftrightarrow \text{$|$-1}\rangle _e\text{$|$0}\rangle_n$ (dashed-dotted, dark blue), and $\text{$|$0}\rangle _e\text{$|$-1}\rangle_n \leftrightarrow \text{$|$-1}\rangle _e\text{$|$-1}\rangle_n$ (dashed, dark blue) are approximately $(2\pi)2856.5$ MHz, $(2\pi)2858.7$ MHz and $(2\pi)2860.8$ MHz, respectively. As expected, the frequency difference between the middle and side dips is $|A_{||}|=(2\pi)2.166\,$ MHz. 

\begin{figure}
    \centering
    \includegraphics[scale=1]{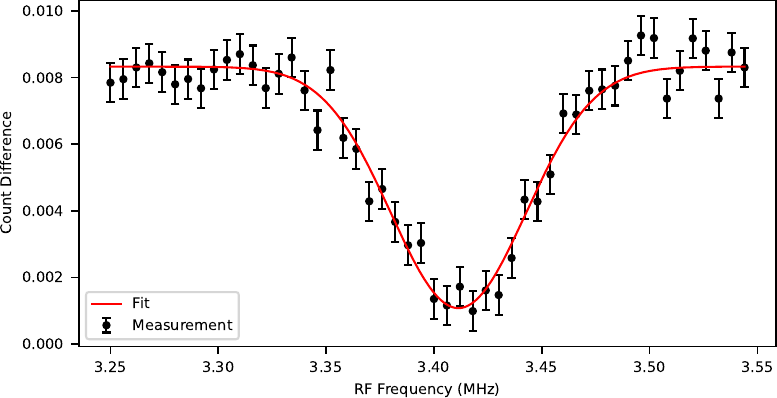}
    \caption{Nuclear ODMR spectrum.
    The data was acquired by probing the transfer of $\mione$ to $\mizero$ by sandwitching an RF $\pi$ pulse with variable frequency between two selective MW $\pi$ pulses.
    On RF resonance at $\rffreq$ the NV center fluorescence signal drops, which was retrieved by fitting the data with \autoref{eq:triple_gaussian} with only a single Gaussian ($i=1$).
    The measurement was acquired for $\sim \SI{15}{\minute}$ with $219510$ measurement sweep repetitions.}
    \label{fig:nuclear_odmr}
\end{figure}

In our experiment, we apply selective radio frequency (RF) pulses, marked with the solid red line in \autoref{fig:SM_scheme} (a), between states $\text{$|$-1}\rangle _e\text{$|$0}\rangle_n \leftrightarrow \text{$|$-1}\rangle _e\text{$|$+1}\rangle_n$. The corresponding RF transition frequency is given by  $|P-A_{||}-\gamma_{\text{N}}B_z|\approx (2\pi)3.41\,$MHz, which is also confirmed experimentally, as shown in \autoref{fig:nuclear_odmr} (b). 

\subsection{System Inialization}\label{SM_Subsection:Initialization}

We describe the initialization protocol of the system. 
We restrict our analysis to $m_\text{s}=-1, 0$ states only as state $m_s=+1$ is highly detuned and we do not address it with our microwave (MW) pulses. 

\emph{Step 1:~~} First, we apply optical pumping to our system by applying a green laser pulse, which ideally initializes the system in the state $\text{$|$0}\rangle _e$. 
The $^{14}$N nuclear spin state population is distributed between states $m_\text{I}=-1, 0, +1$. The populations of the nuclear spin states after long laser illumination can be inferred from the ODMR spectrum in \autoref{fig:SM_scheme} (b) and are estimated from the fit of \autoref{eq:triple_gaussian} as $(\nuclearpopbeforeone,\nuclearpopbeforezero,\nuclearpopbeforemone)$ for $m_\text{I}=+1, 0, -1$, respectively, with the corresponding drops in fluorescence denoted from left to right (the $m_\text{I}=+1$ population is estimated from the area under the left one). 
These values are taken as an example for visualizing the polarization process of the protocol.
Thus, the populations after the green laser pulse are expected to be $P_{\text{$|$0}\rangle _e\text{$|$+1}\rangle_n}=\densitybeforeone$, $P_{\text{$|$0}\rangle _e\text{$|$0}\rangle_n}=\densitybeforezero$, and $P_{\text{$|$0}\rangle _e\text{$|$-1}\rangle_n}=\densitybeforemone$. 

\emph{Step 2:~~} 
We apply a selective MW $\pi$ pulse on the transition $\text{$|$0}\rangle _e\text{$|$+1}\rangle_n \leftrightarrow \text{$|$-1}\rangle _e\text{$|$+1}\rangle_n$, which exchanges the populations of the two states. It acts as a $\cnotn$ pulse, exchanging the populations of the electron spin states $m_s=0$ and $m_s=-1$ if the nuclear spin state is $m_I=+1$. The populations after the selective MW pulse are $P_{\text{$|$-1}\rangle _e\text{$|$+1}\rangle_n}=\densitybeforeone$, $P_{\text{$|$0}\rangle _e\text{$|$0}\rangle_n}=\densitybeforezero$, and $P_{\text{$|$0}\rangle _e\text{$|$-1}\rangle_n}=\densitybeforemone$. 

\emph{Step 3:~~} 
We apply a selective RF $\pi$ pulse on the transition $\text{$|$-1}\rangle _e\text{$|$0}\rangle_n \leftrightarrow \text{$|$-1}\rangle _e\text{$|$+1}\rangle_n$, which exchanges the populations of the two states. 
It acts as a $\cnote$ pulse, exchanging the populations of the nuclear spin state $m_I=+1$ and $m_I=0$ if the electron spin state is $m_s=-1$. 
The populations after the selective RF pulse are $P_{\text{$|$-1}\rangle _e\text{$|$}0\rangle_n}=\densitybeforeone$, $P_{\text{$|$0}\rangle _e\text{$|$+1}\rangle_n}=\densitybeforezero$, and $P_{\text{$|$0}\rangle _e\text{$|$-1}\rangle _n}=\densitybeforemone$. 
We can also consider the populations of the electron and nuclear spin qutrits in the single qutrit bases. 
Then, the populations of the electron spin states are $P_{\text{$|$-1}\rangle _e}=\densitybeforeone$ and $P_{\text{$|$0}\rangle _e}=\densitybeforezeroplusmone$ and the populations of the nuclear $^{14}$N spin states are $P_{\text{$|$0}\rangle _n}=\densitybeforenuclearzero$ and $P_{\text{$|$-1}\rangle _n}=\densitybeforenuclearmone$.

\emph{Step 4:~~} We apply optical pumping to our system by applying a green laser pulse, which ideally initializes the electron spin in state $\text{$|$0}\rangle _e$ without affecting the populations of the nuclear spin state. 
Thus, the populations after the second green laser pulse are expected to be $P_{\text{$|$-1}\rangle _e}=1$, $P_{\text{$|\pm$1}\rangle _e}=0$ and the populations of the nuclear $^{14}$N spin states are preserved $P_{\text{$|$0}\rangle _n}=\densitybeforenuclearzero$ and $P_{\text{$|$-1}\rangle _n}=\densitybeforenuclearmone$. 
The populations in the two-qutrit basis are then $P_{\text{$|$0}\rangle _e\text{$|$0}\rangle}=\densitybeforenuclearzero$, and $P_{\text{$|$0}\rangle _e\text{$|$-1}\rangle}=\densitybeforenuclearmone$. 

\begin{figure}
    \centering
    \includegraphics[scale=1]{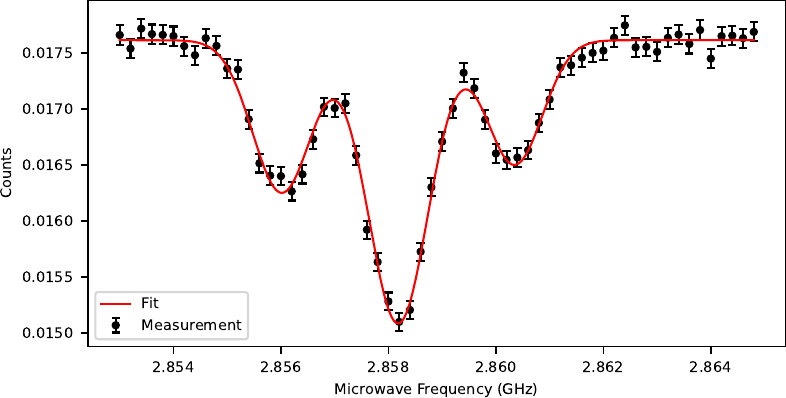}
    \caption{Pulsed ODMR after performing the initialization step of the protocol in \autoref{fig:SM_scheme}.
    After initialization $\nuclearpopafterzero$ of the population is in $\mizero$.
    However, a significant $\nuclearpopafterone$ remains in $\mione$ due to imperfect laser initialization of the NV center.
    The measurement was acquired for $\sim \SI{110}{\minute}$ with $2424256$ measurement sweep repetitions.}
    \label{fig:nuclear_init}
\end{figure}

As a result of the application of steps 2-4, the population of the $m_I=+1$ state is ideally pumped to $m_I=0$, i.e., the left ODMR peak from \autoref{fig:SM_scheme} (b) is transferred to the middle peak. 
During the experimental implementation, we noted that the initialization fidelity of the electron spin with our laser pulse is imperfect, so the population is not entirely pumped to $m_I=0$, as depicted in \autoref{fig:nuclear_init}.
Nevertheless, this was not crucial for our proof-of-principle sensing experiment, as the imperfect initialization only led to a reduced overall contrast. 

The procedure in steps 2-4 can in principle be repeated in a modified version to pump the population of the $m_I=-1$ state to $m_I=0$ (not shown in \autoref{fig:SM_scheme}). 
This can be achieved by applying a selective MW $\pi$ pulse on the transition $\text{$|$0}\rangle _e\text{$|$-1}\rangle_n \leftrightarrow \text{$|$-1}\rangle _e\text{$|$-1}\rangle_n$, which exchanges the population of the two states. 
Then, we can apply a selective RF $\pi$ pulse on the $\text{$|$-1}\rangle _e\text{$|$-1}\rangle_n \leftrightarrow \text{$|$-1}\rangle _e\text{$|$0}\rangle_n$ to exchange the population of the two states, resulting in $P_{\text{$|$0}\rangle _e\text{$|$0}\rangle}=\densitybeforenuclearzero$, and $P_{\text{$|$-1}\rangle _e\text{$|$0}\rangle_n}=\densitybeforenuclearmone$. A subsequent laser pulse would then ideally pump all the population to $m_s=0$, resulting in $P_{\text{$|$0}\rangle _e\text{$|$0}\rangle_n}=1$. In other words, the right ODMR peak from \autoref{fig:SM_scheme} (b) can also be pumped to the middle peak. 
Due to the imperfect initialization fidelity of the electron spin with our laser pulse and the small initial population of $m_I=-1$, we did not apply this initialization step. 
Again, this was not crucial for our proof-of-principle sensing experiment, as the imperfect initialization only led to a reduced overall contrast. 

Polarization of the nitrogen nuclear spin was determined using an ODMR spectrum of the NV center.
A weak MW pulse is used to resolve the hyperfine structure and the data is fit with
\begin{equation}\label{eq:triple_gaussian}
    f(x) = b + \sum_{i=1}^3 a_i \exp \left( - \frac{\left(x - c_i\right)^2}{2 \sigma_i^2} \right)\,,
\end{equation}
where $a_i$ are the amplitudes, $c_i$ the center frequencies, $\sigma_i$ the standard deviation of the three Gaussian distributions and $b$ is a common offset.

When comparing the polarization from \autoref{fig:SM_scheme} (b) to \autoref{fig:nuclear_init} one can see that the nuclear spin polarization reduces from not performing the initialization step ($\nuclearpopbeforeone$ population in $m_I = +1$) to performing it ($\nuclearpopafterzero$ population in $m_I = 0$).
Following an extensive investigation, we have identified laser leakage as the main reason for the limited initialization fidelity. Specifically, some laser radiation continues to reach the sample during time periods when the laser should be turned off.
This causes a slow, constant repumping of the NV center during the CNOT gates of the initialization sequence, leading to limited initialization fidelity.

The preparation step with repumping to the middle state is necessary to ensure the same initial polarization of the nuclear spin.
Specifically, the polarization shown in \autoref{fig:SM_scheme} (b) is not intrinsically available in our system.
If the system is left untouched for longer than the nuclear spin lifetime $\widetilde{T}_{1, \mathrm{nuc}}$, there will be almost no polarization of the nuclear spin. 
The polarization of the nuclear spin in \autoref{fig:SM_scheme} is built up by slow flip-flop processes \cite{jacques2009dynamic} during laser illumination.
The ODMR spectrum is acquired continuously, which means that the microwave pulse's (duration: $\sim \SI{1.5}{\micro \second}$) frequency is swept, followed by a $\SI{3}{\micro \second}$ laser.
Once data for all frequency points have been acquired, the whole sequence is repeated for $\sim \SI{20}{\minute}$.
This means that most of the measurement time is laser illumination, resulting in the slow flip-flop processes described in \cite{jacques2009dynamic} taking place.
The visible polarization is the result of many of these flip-flop cycles, each building up a little bit of polarization and with no waiting time in between to let the nuclear spin decay again.
In order to create the nuclear spin polarization without an initialization sequence, one would need a laser pulse, longer than the nuclear $\widetilde{T}_{1, \mathrm{nuc}}$.
From \autoref{fig:nuclear_init_decay} one can see that this would require a laser pulse longer than $\SI{210}{\micro \second}$, which is significantly longer than running our initialization sequence ($\sim \SI{3}{\micro \second}$ laser + $\SI{1}{\micro \second}$ waiting time + $\SI{1.5}{\micro \second}$ MW pulse + $\SI{41}{\micro \second}$ RF pulse $\approx \SI{46.5}{\micro \second}$). 
In summary, this analysis shows that the initial nuclear spin polarization would vary without the repumping step and would depend on the laser illumination time, which is typically shorter in step III of the MCS protocol than in step I to optimize $\widetilde{T}_{1, \mathrm{nuc}}$. The preparation step with repumping to the middle state is then necessary to avoid these drawbacks and ideally ensure the same initial polarization of the nuclear spin.

The difference in population of the $m_I = -1$ state in \autoref{fig:SM_scheme} (b) and \autoref{fig:nuclear_init} is due to the measurements being taken 3 days apart.
During this time, the setup can experience thermal drift, causing the left resonance line to drift from $\SI{2.856526 +- 0.000009}{\giga \hertz}$ in \autoref{fig:SM_scheme} (b) to $\SI{2.856015 +- 0.000012}{\giga \hertz}$ in \autoref{fig:nuclear_init}.
Changes in the magnetic field can cause a change in the flip-flop dynamics \cite{jacques2009dynamic}, resulting in different polarization of the nuclear spin under long laser illumination.

\subsection{System Hamiltonians and Propagators}\label{SM_Subsection:Pulses_Theory}

We describe in detail the Hamiltonians and the propagators that characterize the system evolution during the MW and RF pulses, which we use in MCS to detect a signal originating from a classical field or from statistical polarization in nano-NMR experiments. 
The total Hamiltonian that takes into account the application of MW and RF driving pulses is given by
\begin{align}\label{eq:SM_Hamiltonian_total}
H_{\text{total}}=H_{^{14}\text{NV-}}+\underbrace{\Omega_{\text{MW}}\sqrt{2}\cos{\left(\omega_{\text{MW}}t+\xi_{\text{MW}}\right)}S_x}_{H_{\text{MW}}}+\underbrace{\Omega_{\text{RF}}\sqrt{2}\cos{\left(\omega_{\text{RF}}t+\xi_{\text{RF}}\right)}I_x}_{H_{\text{RF}}}+\underbrace{g_{\text{s}}\cos{\left(\omega_{\text{s}}t+\xi_{\text{s}}\right)}S_z}_{H_{\text{signal}}},
\end{align}
where $\Omega_{\text{MW}}$, $\omega_{\text{MW}}$ and $\xi_{\text{MW}}$ are the Rabi frequency, carrier frequency and the initial phase, respectively of the applied MW field; $\Omega_{\text{RF}}$, $\omega_{\text{RF}}$ and $\xi_{\text{RF}}$ are the respective values for the RF field, and $g_{\text{s}}$, $\omega_{\text{s}}$ and $\xi_{\text{s}}$ are the respective values for the signal field we want to sense. We note that the signal field could arise from a classical signal, which is not phase locked with the MW and RF control fields we apply, i.e., $\xi_{\text{s}}$ varies from run to run. Alternatively, the signal could be due to statistical polarization in the case of nano-NMR experiments where both $g_{\text{s}}$ and $\xi_{\text{s}}$ can vary from run to run or for different NV centers in an ensemble. 
It proves useful to move to the rotating frame, defined by $H_{\text{rot}}=\omega_{\text{MW}}S_z+\omega_{\text{RF}}I_z+\left(A_{||}-2P/3\right)1\otimes 1$. The resulting interaction Hamiltonian takes the form
\begin{equation}
    H_{\text{int}}=U_{\text{rot}}(t)H_{\text{total}}U_{\text{rot}}(t)^{\dagger}-i U_{\text{rot}}(t)\partial_t U_{\text{rot}}(t)^{\dagger},\qquad U_{\text{rot}}(t)=\exp{\left(-i H_{\text{rot}}t\right)}.
\end{equation}
We work in the subspace of states $\{\text{$|$0}\rangle _e\text{$|$0}\rangle_n, \text{$|$0}\rangle _e\text{$|$+1}\rangle_n, \text{$|$-1}\rangle _e\text{$|$0}\rangle_n, \text{$|$-1}\rangle_e\text{$|$+1}\rangle_n\}$, which we label $\{|1\rangle, |2\rangle,|3\rangle,|4\rangle \}$, respectively. The resulting reduced interaction Hamiltonian for the dynamics within these four states, using that $\omega_{\text{MW}}=\gamma _{\text{NV}}B_z-\text{DD}+A_{||}-\delta _{\text{MW}}$ and $\omega_{\text{RF}}=P-A_{||}-\gamma _{\text{N14}}B_z-\delta _{\text{RF}}$, where $\delta _{\text{MW}}$ ($\delta _{\text{RF}}$) is the detuning of the respective frequency from the MW (RF) driving field, after applying the rotating-wave approximation (RWA) and removing the fast oscillating terms by using that $\Omega_{\text{MW}}\ll \omega_{\text{MW}}$, $\Omega_{\text{RF}}\ll \omega_{\text{RF}}, A_{||}, P$ takes the form 
\begin{equation}
    H_{\text{int,rwa}}\approx \left[
\begin{array}{cccc}
 -\text{$A_{||}$} & \frac{1}{2} e^{i \xi _{\text{RF}}} \Omega _{\text{RF}} & \frac{1}{2} e^{-i \xi _{\text{MW}}} \Omega
   _{\text{MW}} & 0 \\
 \frac{1}{2} e^{-i \xi _{\text{RF}}} \Omega _{\text{RF}} & \delta _{\text{RF}} & 0 & \frac{1}{2} e^{-i \xi _{\text{MW}}}
   \Omega _{\text{MW}} \\
 \frac{1}{2} e^{i \xi _{\text{MW}}} \Omega _{\text{MW}} & 0 & -\delta _{\text{MW}} & \frac{1}{2} e^{i \xi _{\text{RF}}}
   \Omega _{\text{RF}} \\
 0 & \frac{1}{2} e^{i \xi _{\text{MW}}} \Omega _{\text{MW}} & \frac{1}{2} e^{-i \xi _{\text{RF}}} \Omega _{\text{RF}} &
   \delta _{\text{RF}}-\delta _{\text{MW}} \\
\end{array}
\right].
\end{equation}
In the following analysis, we assume that $\delta _{\text{MW}}\ll\Omega_{\text{MW}}$ ($\delta _{\text{RF}}\ll\Omega_{\text{RF}}$), so we can neglect them during the respective pulses.  

First, we consider the Hamiltonian when we use strong MW pulses to drive the electron spin, independently of the nuclear spin state. The Hamiltonian in the rotating frame after applying the RWA takes the form: 
\begin{align}
    H_{\text{MW,2q}}\approx\Omega_{\text{MW}} \left(\cos{(\xi_{\text{MW}})} s_x + \sin{(\xi_{\text{MW}})} s_y\right) &=\left[
\begin{array}{cccc}
 0 & 0 & \frac{\Omega_{\text{MW}}}{2} e^{-i \xi_{\text{MW}} }  & 0 \\
 0 & 0 & 0 & \frac{\Omega_{\text{MW}}}{2} e^{-i \xi_{\text{MW}} }  \\
 \frac{\Omega_{\text{MW}}}{2} e^{i \xi_{\text{MW}} }  & 0 & 0 & 0 \\
 0 & \frac{\Omega_{\text{MW}}}{2} e^{i \xi_{\text{MW}} }  & 0 & 0 \\
\end{array}
\right],
\end{align}
where $\xi_{\text{MW}}$ is the phase of the respective pulse, e.g., $\xi_{\text{MW}}=0^{\circ}$ for an $X$ pulse and $\xi_{\text{MW}}=90^{\circ}$ for a $Y$ pulse, $s_{i}=\sigma_i/2$ are spin $1/2$ operators, where $\sigma_i=x,y$ are the respective Pauli operators. We assumed that the driving is strong and we can neglect the effect of detunings and phase evolution due to $A_{||}$ during the pulse. The evolution of the system due to the pulse is characterized by the respective propagators in the two-qubit basis and single qubit basis of the electron spin. 
\begin{align}
    U_{\text{MW,2q}}(A)=\exp{\left(-i H_{\text{MW,2q}} t\right)}&=\left[
\begin{array}{cccc}
 \cos \left(\frac{A}{2}\right) & 0 & -i e^{-i \xi_{\text{MW}} } \sin \left(\frac{A}{2}\right) & 0 \\
 0 & \cos \left(\frac{A}{2}\right) & 0 & -i e^{-i \xi_{\text{MW}} } \sin \left(\frac{A}{2}\right) \\
 -i e^{i \xi_{\text{MW}} } \sin \left(\frac{A}{2}\right) & 0 & \cos \left(\frac{A}{2}\right) & 0 \\
 0 & -i e^{i \xi_{\text{MW}} } \sin \left(\frac{A}{2}\right) & 0 & \cos \left(\frac{A}{2}\right) \\
\end{array}
\right],\\
U_{\text{MW,e}}(A)&=\left[
\begin{array}{cc}
 \cos \left(\frac{A}{2}\right) & -i e^{-i \xi_{\text{MW}} } \sin \left(\frac{A}{2}\right) \\
 -i e^{i \xi_{\text{MW}} } \sin \left(\frac{A}{2}\right) & \cos \left(\frac{A}{2}\right) \\
\end{array}
\right],
\end{align}
where $A=\Omega_{\text{MW}} t$ is the pulse area, which can be e.g., $\pi/2$ or $\pi$. 

In addition to the strong MW pulse, we also apply a selective MW $\pi$ pulse on the transition $|2\rangle\leftrightarrow |4\rangle$, i.e., $\text{$|$0}\rangle _e\text{$|$+1}\rangle_n \leftrightarrow \text{$|$-1}\rangle _e\text{$|$+1}\rangle_n$, which exchanges the populations of the two states. 
It acts as a $\cnotn$ gate, apart from a phase shift, exchanging the populations of the electron spin states $m_s=0$ and $m_s=-1$ if the nuclear spin state is $m_I=+1$. 
The Hamiltonian in the two-qubit basis during the pulse in the rotating frame after applying the RWA and the respective propagator takes the form:
\begin{align}
    H_{\text{\cnotn}} &=\left[
\begin{array}{cccc}
 0 & 0 & 0 & 0 \\
 0 & 0 & 0 & \frac{g_{\text{MW}}}{2} \\
 0 & 0 & 0 & 0 \\
 0 & \frac{g_{\text{MW}}}{2} & 0 & 0 \\
\end{array}
\right],\qquad
U_{\text{\cnotn}}=\exp{\left(-i H_{\text{\cnotn}} t_{\text{\cnotn}}\right)}\left[
\begin{array}{cccc}
 1 & 0 & 0 & 0 \\
 0 & 0 & 0 & -i \\
 0 & 0 & 1 & 0 \\
 0 & -i & 0 & 0 \\
\end{array}
\right],
\end{align}
where $g_{\text{MW}}$ is the weak Rabi frquency of the selective MW pulse, so that it addresses only the $\text{$|$0}\rangle _e\text{$|$+1}\rangle_n \leftrightarrow \text{$|$-1}\rangle _e\text{$|$+1}\rangle_n$ transition. The duration of the $\cnotn$ pulse is chosen, so that its pulse area is $g_{\text{MW}} t_{\text{\cnotn}}=\pi$ and it exchanges the populations of states $|2\rangle$ and $|4\rangle$, i.e., $\text{$|$0}\rangle _e\text{$|$+1}\rangle_n \leftrightarrow \text{$|$-1}\rangle _e\text{$|$+1}\rangle_n$. We note that the propagator $U_{\text{\cnotn}}$ does not only exchange states $|2\rangle$ and $|4\rangle$ as a typical $\cnotn$ gate would do, but gives an additional prefactor $-i$ to the probability amplitudes of both states, but this is irrelevant for our sensing scheme.  

Finally, our protocol also contains a selective RF $\pi$ pulse on the transition $|3\rangle\leftrightarrow |4\rangle$, i.e., $\text{$|$-1}\rangle _e\text{$|$0}\rangle_n \leftrightarrow \text{$|$-1}\rangle _e\text{$|$+1}\rangle_n$, which exchanges the populations of the two states. It acts as a $\cnote$ gate, apart from a phase shift, exchanging the nuclear spin states $m_I=+1$ and $m_I=0$ if the electron spin state is $m_s=-1$. 
The Hamiltonian in the two-qubit basis during the pulse in the rotating frame after applying the RWA and the respective propagator take the form:
\begin{align}
    H_{\text{\cnote}} &=\left[
\begin{array}{cccc}
 0 & 0 & 0 & 0 \\
 0 & 0 & 0 & 0 \\
 0 & 0 & 0 & \frac{g_{\text{RF}}}{2} \\
 0 & 0 & \frac{g_{\text{RF}}}{2} & 0 \\
\end{array}
\right],\qquad
U_{\text{\cnote}}=\exp{\left(-i H_{\text{\cnote}} t_{\text{\cnote}}\right)}\left[
\begin{array}{cccc}
 1 & 0 & 0 & 0 \\
 0 & 1 & 0 & 0 \\
 0 & 0 & 0 & -i \\
 0 & 0 & -i & 0 \\
\end{array}
\right],
\end{align}
where $g_{\text{RF}}$ is the weak Rabi frquency of the selective RF pulse. 
The duration of the $\cnote$ pulse is chosen, so that its pulse area is $g_{\text{RF}} t_{\text{\cnote}}=\pi$ and it exchanges states $|3\rangle$ and $|4\rangle$. 
Again, we note that the propagator $U_{\text{\cnote}}$ adds an additional prefactor $-i$ to the probability amplitudes of both states, which are exchanged in contrast to a typical CNOT gate, but this is not important for our sensing scheme. Finally, we note that, for simplicity of presentation and without loss of generality, we have neglected phase evolution on the electron and nuclear spin states due to unwanted detunings during the RF pulse.

\subsection{Detailed Theory of Multipoint Correlation Spectroscopy}\label{SM_Subsection:MCS_Theory}

We describe in detail the theory of the multipoint correlation spectroscopy protocol. We note that the description is general as the protocol itself is applicable to any two qubits where one acts as a sensor and the second one as a memory qubit. We work in the subspace of states $\{\text{$|$0}\rangle _e\text{$|$0}\rangle_n, \text{$|$0}\rangle _e\text{$|$+1}\rangle_n, \text{$|$-1}\rangle _e\text{$|$0}\rangle_n, \text{$|$-1}\rangle_e\text{$|$+1}\rangle_n\}$, which we label $\{|1\rangle, |2\rangle,|3\rangle,|4\rangle \}$, respectively. 

\subsection*{Step-by-step description of MCS}

\emph{Step I: Initialization of the NV center electron spin and 14N nuclear spin.~}

We assume for simplicity of presentation and without loss of generality that the system is initialized in state $|1\rangle$.  
Imperfect initialization would usually lead to reduced contrast, but would not significantly affect the performance of the protocol as described in \autoref{SM_Subsection:Initialization}. 
The density matrix after the initialization is then given by 
\begin{equation}
    \rho_{\text{in,e,n}}\approx  \text{$|$0}\rangle _e\text{$|$0}\rangle_n \langle \text{0$|$}_e\langle\text{0$|$}_n =\left[
\begin{array}{cccc}
 1 & 0 & 0 & 0 \\
 0 & 0 & 0 & 0 \\
 0 & 0 & 0 & 0 \\
 0 & 0 & 0 & 0 \\
\end{array}
\right],~
\rho_{\text{in,e}}\approx  \text{$|$0}\rangle _e\langle \text{0$|$}_e=\left[
\begin{array}{cc}
 1 & 0  \\
 0 & 0 
\end{array}
\right],~
\rho_{\text{in,n}}\approx  \text{$|$0}\rangle _n\langle \text{0$|$}_n=\left[
\begin{array}{cc}
 1 & 0  \\
 0 & 0 
\end{array}
\right],
\end{equation}
where the first density matrix is in the two-qubit basis, while the second and third ones are in the reduced single qubit bases of the NV center electron spin, comprising states $\{\text{$|$0}\rangle _e,\text{$|$-1}\rangle _e\}$, and the $^{14}$N nuclear spin, which consists of $\{\text{$|$0}\rangle_n, \text{$|$+1}\rangle_n\}$. 

\emph{Step II: Initial phase acquisition and storage of phase information in the populations of the memory qubit.~}

During the initial phase accumulation period of the experiment, we accumulate a phase which is subsequently stored as a population difference in the $^{14}$N nuclear spin memory qubit. Specifically, we first apply a $\pi/2$ strong MW pulse, which creates a coherent superposition state between states $\text{$|$0}\rangle_e$ and $\text{$|$-1}\rangle_e$, so that
\begin{equation}
    \rho_{\text{II,1,e,n}} =\left[
\begin{array}{cccc}
 \frac{1}{2} & 0 & \frac{i}{2} & 0 \\
 0 & 0 & 0 & 0 \\
 -\frac{i}{2} & 0 & \frac{1}{2} & 0 \\
 0 & 0 & 0 & 0 \\
\end{array}
\right],~
\rho_{\text{II,1,e}}=\left[
\begin{array}{cc}
 \frac{1}{2} & \frac{i}{2} \\[1pt]
 -\frac{i}{2} & \frac{1}{2} \\
\end{array}
\right],~
\rho_{\text{II,1,n}} =\left[
\begin{array}{cc}
 1 & 0  \\
 0 & 0 
\end{array}
\right].
\end{equation}

We then apply dynamical decoupling (DD) with an XY8 sequence \cite{Gullion1990jmr} during the phase eqcuisition time.
We note that DD is not crucial for the protocol and phase accumulation as MCS can also be applied without it. However, then the phase accumulation duration without DD is limited by $T_2^*$, which can be quite low, especially in spin ensembles, e.g., of the order of tens of nanoseconds for ensembles NV electron spins \cite{GenovPRR2020}. DD acts as a noise filter and prolongs the coherence time by several orders of magnitude, e.g., to $T_2$ of the order of milliseconds even with NV electron spin ensembles \cite{GenovPRR2020}. Therefore, DD is especially useful for detecting \emph{weak oscillating signals} \cite{DegenRMP2017}.  DD sequences of equal pulse separation, which we use, typically perform better for common noise spectra without a sharp frequency cutoff \cite{Suter2016RevModPhys,DegenRMP2017}. Their coherence time is an important factor, which typically depends on the sequences' robustness to pulse errors \cite{RDD_review12Suter,Suter2016RevModPhys}. 
We apply the XY8 sequence on resonance for optimal phase acquisition \cite{DegenRMP2017}, i.e., we choose 
the time period between the centers of the strong MW $\pi$ pulses is $\tau\approx 1/(2\nu_s)$, where $\nu_s$ is the frequency of the signal we try to detect with $\omega_s=(2\pi)\nu_s$ -- the respective angular frequency, in case we perform quantum sensing of an artificial signal or the Larmor frequency of the spins for nano-NMR with statistical polarization. The accumulated phase during this initial run is given by \cite{DegenRMP2017}
\begin{equation}
\phi_0^{(j)}\approx\frac{2}{\pi}g^{(j)}_{s,0}t_{\text{DD}}\cos{\left(\xi_{s,0}^{(j)}\right)}=\frac{2}{\pi}\gamma_{\text{NV}}B_{s,k}^{(j)}t_{\text{DD}}\cos{\left(\xi_{s,0}^{(j)}\right)},
\end{equation}
where $g_{s,0}^{(j)}$ is the signal amplitude during the initial phase accumulation period for the $j$-th NV center in an ensemble or the $j$-th independent experimental run, which is proportional to the signal magnetic field envelope $B_{s,0}^{(j)}$ that we assume to be constant during the XY8 sequence, which is feasible in our experiment.
In addition,  $t_{\text{DD}}$ is the duration of the phase acquisition period, i.e., the length of the XY8 sequence, and $\xi_{s,0}^{(j)}$ is the initial phase of the signal at its start. The density matrix of the system takes the form 
\begin{equation}
    \rho_{\text{II,2,e,n}}^{(j)} =\left[
\begin{array}{cccc}
 \frac{1}{2} & 0 & \frac{1}{2} i e^{-i \phi _0^{(j)}} & 0 \\
 0 & 0 & 0 & 0 \\
 -\frac{1}{2} i e^{i \phi _0^{(j)}} & 0 & \frac{1}{2} & 0 \\
 0 & 0 & 0 & 0 \\
\end{array}
\right],~
\rho_{\text{II,2,e}}^{(j)}=\left[
\begin{array}{cc}
 \frac{1}{2} & \frac{1}{2} i e^{-i \phi _0^{(j)}} \\
 -\frac{1}{2} i e^{i \phi _0^{(j)}} & \frac{1}{2} \\
\end{array}
\right],~
\rho_{\text{II,2,n}}^{(j)} =\left[
\begin{array}{cc}
 1 & 0  \\
 0 & 0 
\end{array}
\right].
\end{equation}

The phase information is then stored in a population difference between the electron spin states by a $\pi/2$ pulse, which is phase shifted in comparison to the initial pulse by $90^{\circ}$, i.e., a $\pi/2$ Y pulse. The density matrix of the system is then given by
\begin{equation}
    \rho_{\text{II,3,e,n}}^{(j)} =\frac{1}{2}\left[
\begin{array}{cccc}
  1-\sin (\phi^{(j)} _0) & 0 &i \cos
   (\phi^{(j)} _0) & 0 \\
 0 & 0 & 0 & 0 \\
 -i \cos (\phi^{(j)} _0) & 0 & 1+\sin (\phi^{(j)}
   _0) & 0 \\
 0 & 0 & 0 & 0 \\
\end{array}
\right],~
\rho_{\text{II,3,e}}^{(j)}=\frac{1}{2}\left[
\begin{array}{cc}
  1-\sin (\phi^{(j)} _0) & i \cos (\phi^{(j)}
   _0) \\
 -i \cos (\phi^{(j)} _0) & 1+\sin (\phi^{(j)}
   _0) \\
\end{array}
\right],~
\rho_{\text{II,3,n}}^{(j)} =\left[
\begin{array}{cc}
 1 & 0  \\
 0 & 0 
\end{array}
\right].
\end{equation}
Information about $\phi _0^{(j)}$ is subsequently stored as a population difference of the nuclear spin (memory qubit) by applying a selective RF pulse, which performs an effective $\cnote$ gate, apart from an unimportant phase shift. The final density matrix after the $\pi/2$ nuclear spin states takes the form: 
\begin{align}
    \rho_{\text{II,4,e,n}}^{(j)} &=\frac{1}{2}\left[
\begin{array}{cccc}
 1-\sin (\phi^{(j)} _0) & 0 & 0 & - \cos
   (\phi^{(j)} _0) \\
 0 & 0 & 0 & 0 \\
 0 & 0 & 0 & 0 \\
 -\cos (\phi^{(j)} _0) & 0 & 0 &  1+\sin (\phi^{(j)}
   _0) \\
\end{array}
\right],~\notag\\
\rho_{\text{II,4,e}}^{(j)}&=\frac{1}{2}\left[
\begin{array}{cc}
 1-\sin (\phi^{(j)} _0) & 0 \\
 0 & 1+\sin (\phi^{(j)} _0) \\
\end{array}
\right],~
\rho_{\text{II,4,n}}^{(j)} =\frac{1}{2}\left[
\begin{array}{cc}
  1-\sin (\phi^{(j)} _0) & 0 \\
 0 &  1+\sin (\phi^{(j)} _0) \\
\end{array}
\right].
\end{align}

\emph{Step III: Subsequent phase acquisitions and retrieval of phase information from the memory qubit to obtain information about phase correlations.~}

We perform a series of subsequent phase accumulation periods by applying the same sequence apart from the final $\cnote$ gate, which we replace with a $\cnotn$ pulse. 
In each experiment, we first apply a laser pulse to prepare the electron spin state in state $|0\rangle_e$. 
We assume that the initialization of the electron spin ideally does not affect the populations of the nuclear spin state. 
This should be feasible when we work at high magnetic fields, as in our experiments. The density matrix after reinitialization during the $k$-th subsequent measurement, $k=1,\dots, M$, takes the form
\begin{equation}
    \rho_{\text{III,1,e,n}}^{(j)}\approx  \frac{1}{2}\left[
\begin{array}{cccc}
  1-\sin (\phi^{(j)} _0) & 0 & 0 & 0 \\
 0 &  1+\sin{(\phi^{(j)} _0)} & 0 & 0 \\
 0 & 0 & 0 & 0 \\
 0 & 0 & 0 & 0 \\
\end{array}
\right],~
\rho_{\text{III,1,e}}^{(j)}\approx  \left[
\begin{array}{cc}
 1 & 0  \\
 0 & 0 
\end{array}
\right],~
\rho_{\text{III,1,n}}^{(j)}\approx  \frac{1}{2}\left[
\begin{array}{cc}
  1-\sin (\phi^{(j)} _0) & 0 \\
 0 &  1+\sin (\phi^{(j)} _0) \\
\end{array}
\right],
\end{equation}

After, the electron spin is initialized, we repeat the experiment in the previous step, where we prepare the electron spin qubit in an equal coherent superposition state by a strong MW $\pi/2$ pulse, apply dynamical decoupling with the XY8 sequence and time between the centers of the strong MW $\pi$ pulses of $\tau\approx 1/(2\nu_s)$, so the NV center electron spin (our sensor qubit) accumulates a phase
\begin{equation}
\phi_k^{(j)}\approx\frac{2}{\pi}g^{(j)}_{s,k}t_{\text{DD}}\cos{\left(\xi_{s,k}^{(j)}\right)}=\frac{2}{\pi}\gamma_{\text{NV}}B_{s,k}^{(j)}t_{\text{DD}}\cos{\left(\xi_{s,0}^{(j)}+\omega_s \Delta T_k\right)},
\end{equation}
where $g_{s,k}^{(j)}$ is the signal amplitude during the $k$-th phase accumulation period for the $j$-th NV center in an ensemble or for the $j$-th repetition of the experiment, and it is proportional to the signal magnetic field envelope $B_{s,k}^{(j)}$, which we assume to be constant during the XY8 sequence; $\xi_{s,k}^{(j)}=\xi_{s,0}^{(j)}+\omega_s \Delta T_k$ is the signal initial phase at the beginning of the $k$-th phase accumulation period. Information about $\phi_k^{(j)}$ is subsequently stored in the population difference between the electron spin states by a strong MW $\pi/2$ Y pulse and the density matrix takes the form
\begin{align}
    &\rho_{\text{III,2,e,n}}^{(j)}=\notag\\ &\frac{1}{4}\left[
\begin{array}{cccc}
 [1-\sin{(\phi^{(j)} _0)}] [1-\sin{(\phi^{(j)} _k)}] & 0 &  i [1-\sin{(\phi^{(j)} _0)}] \cos{(\phi^{(j)} _k)} & 0 \\
 0 & [1+\sin{(\phi^{(j)} _0)}] [1-\sin{(\phi^{(j)} _k)}] & 0 &  i [1+\sin (\phi^{(j)} _0)]
   \cos (\phi^{(j)} _k) \\
 - i [1-\sin (\phi^{(j)} _0)] \cos (\phi^{(j)} _k) & 0
   & [1-\sin (\phi^{(j)} _0)] [1+\sin{(\phi^{(j)} _k)} & 0 \\
 0 & - i [1+\sin (\phi^{(j)} _0)] \cos (\phi^{(j)} _k)
   & 0 &  [1+\sin (\phi^{(j)} _0)] [1+\sin{(\phi^{(j)}_k)}] \\
\end{array}
\right],~\notag\\
&\rho_{\text{III,2,e}}^{(j)}=\frac{1}{2}\left[
\begin{array}{cc}
 1-\sin (\phi^{(j)} _k) &  i \cos (\phi^{(j)}
   _k) \\
 - i \cos (\phi^{(j)} _k) &  1+\sin (\phi^{(j)}
   _k) \\
\end{array}
\right],~
\rho_{\text{III,2,n}}^{(j)} =\frac{1}{2}\left[
\begin{array}{cc}
  1-\sin (\phi^{(j)} _0) & 0 \\
 0 &  1+\sin (\phi^{(j)} _0) \\
\end{array}
\right].
\end{align}

The phase information from the initial phase accumulation period $\phi_0^{(j)}$ is retrieved after every $k$-th phase acquisition with a weak, selective MW pulse, acting as as $\cnotn$ gate, apart from an unimportant phase shift. The density matrix afterwards takes the form
\begin{align}
    &\rho_{\text{III,3,e,n}}^{(j)}=\notag\\ &\frac{1}{4}\left[
\begin{array}{cccc}
 [1-\sin{(\phi_0^{(j)})}] [1-\sin{(\phi_k^{(j)})}] & 0 & i [1-\sin{(\phi_0^{(j)})}] \cos
   {(\phi_k^{(j)})} & 0 \\
 0 & [1+\sin{(\phi_0^{(j)})}] [1+\sin{(\phi_k^{(j)})}] & 0 & -i [1+\sin{(\phi_0^{(j)})}] \cos{(\phi_k^{(j)})} \\
 -i [1-\sin{(\phi_0^{(j)})}] \cos{(\phi_k^{(j)})} & 0 & [1-\sin{(\phi_0^{(j)})}] [1+\sin{(\phi_k^{(j)})}] & 0 \\
 0 & i [1+\sin{(\phi_0^{(j)})}] \cos{(\phi_k^{(j)})} & 0 & [1+\sin{(\phi_0^{(j)})}] [1-\sin{(\phi_k^{(j)})}] \\
\end{array}
\right],~\notag\\
&\rho_{\text{III,3,e}}^{(j)}=\frac{1}{2}\left[
\begin{array}{cc}
 1+\sin {(\phi _0^{(j)})} \sin {(\phi _k^{(j)})} & -i \sin {(\phi _0^{(j)})} \cos {(\phi _k^{(j)})} \\
 i \sin {(\phi _0^{(j)})} \cos {(\phi _k^{(j)})} & 1-\sin {(\phi _0^{(j)})} \sin {(\phi _k^{(j)})} \\
\end{array}
\right],~
\rho_{\text{III,3,n}}^{(j)} =\rho_{\text{II,4,n}}=\frac{1}{2}\left[
\begin{array}{cc}
  1-\sin{(\phi^{(j)} _0)} & 0 \\
 0 &  1+\sin{(\phi^{(j)} _0)} \\
\end{array}
\right].
\end{align}

The expected population of the electron spin state $|0\rangle_e$ is then given by 
\begin{equation}
P_{|0\rangle_e}= \frac{1}{2}\left(1+\langle\sin (\phi^{(j)} _0) \sin (\phi^{(j)} _k) \rangle\right)\approx  \frac{1}{2}+\frac{1}{2}\langle\phi^{(j)} _0\phi^{(j)} _k\rangle,
\end{equation}
where the last approximation is valid for small $\phi^{(j)}_0\ll1$ and $\phi^{(j)}_k\ll1$. The density matrix of the nuclear spin (the storage qubit) is the same as at the end of step II, so the process can be repeated for $k=1,\dots,M$.

\subsection*{Validity of the linear approximation for the observed signal}

We note that the linear approximation $\sin (\phi^{(j)} _0) \sin (\phi^{(j)} _k)\approx \phi^{(j)} _0\phi^{(j)} _k$ is typical for many quantum sensing experiments \cite{DegenRMP2017} but it is not crucial for the MCS protocol. One can understand its theoretical bounds by the following analysis. 
We can perform a Taylor expansion of 
\begin{equation}
\sin{(\phi_k^{(j)})}= \phi_k^{(j)}-\frac{(\phi_k^{(j)})^3}{6}+O\left((\phi_k^{(j)})^5\right).
\end{equation}
Assuming the linear approximation $\sin{(\phi_k^{(j)})}\approx \phi_k^{(j)}$, we obtain 
\begin{align}
\langle\sin{(\phi_0^{(j)})}\sin{(\phi_k^{(j)})}\rangle&\approx                                 \langle\phi_0^{(j)}\phi_k^{(j)}\rangle\notag=
\left\langle\frac{2}{\pi}\gamma_{\text{NV}}B_0^{(j)}t_{\text{DD}}\cos{(\xi_0^{(j)})}\frac{2}{\pi}\gamma_{\text{NV}}B_k^{(j)}t_{\text{DD}}\cos{(\xi_0^{(j)}+\omega_s \Delta T_k)}\right\rangle\notag\\
&=\frac{4}{\pi^2}\gamma_{\text{NV}}^2t_{\text{DD}}^2\left\langle B_0^{(j)}B_k^{(j)}\cos{(\xi_0^{(j)})}\cos{(\xi_0^{(j)}+\omega_s \Delta T_k)}\right\rangle\notag\\
&=\frac{4}{\pi^2}\gamma_{\text{NV}}^2t_{\text{DD}}^2 B_{\text{rms}}^2\cos{(\omega_s \Delta T_k)}\notag\\
&= \phi_{\text{rms}}^2 \cos{\left(\omega_s \Delta T_k \right)},~\phi_{\text{rms}}=\frac{2}{\pi}\gamma_{\text{NV}} B_{\text{rms}} t_{\text{DD}},
\end{align}
where $B_{\text{rms}}^2=B_0^2/2$ when we sense a coherent signal with a magnetic field amplitude $B_0=B_0^{(j)}=B_k^{(j)}$ with a random initial phase $\xi_0^{(j)}$ over which we average. In the case of statistical polarization, $B_{\text{rms}}^2$ is the variance of the magnetic field in the detection volume \cite{DegenRMP2017}.

Next, we take into account the third order correction of the signal, as follows
\begin{align}
\langle\sin{(\phi_0^{(j)})}\sin{(\phi_k^{(j)})}\rangle &\approx \left\langle\left(\phi_0^{(j)}-\frac{(\phi_0^{(j)})^3}{6}\right)\left(\phi_k^{(j)}-\frac{(\phi_k^{(j)})^3}{6}\right)\right\rangle
    = \left\langle\phi_0^{(j)}\phi_k^{(j)}-\frac{(\phi_0^{(j)})^3\phi_k^{(j)}}{6}-\frac{\phi_0^{(j)}(\phi_k^{(j)})^3}{6}+\frac{(\phi_0^{(j)}\phi_k^{(j)})^3}{36}\right\rangle\notag\\
    &\approx \phi_{\text{rms}}^2 \cos{\left(\omega_s \Delta T_k \right)}-\frac{\phi_{\text{rms}}^4}{4} \cos{\left(\omega_s \Delta T_k \right)}+\frac{\phi_{\text{rms}}^6}{72}  \cos \left(\omega_s\Delta T_k\right)+\frac{\phi _{\text{rms}}^6}{288}  \cos \left(\text{$\Delta
   $T}_k \omega _s\right) \cos \left(2 \text{$\Delta $T}_k
   \omega _s\right)\notag\\
   &= \phi_{\text{rms}}^2 \cos{\left(\omega_s \Delta T_k \right)}\left(1-\frac{\phi_{\text{rms}}^2}{4}+\frac{\phi_{\text{rms}}^4}{72}\right)+\frac{\phi _{\text{rms}}^6}{288} \cos{\left(\omega_s \Delta T_k \right)} \cos \left(2 \omega_s\text{$\Delta $T}_k\right).
\end{align}
It is evident that the inclusion of the third order terms in the expansion of $\sin{(\phi_k^{(j)})}$ only leads to a small correction of the amplitude of the signal that oscillates at the angular frequency $\omega_s$ with a factor $1-\frac{\phi_{\text{rms}}^2}{4}+\frac{\phi_{\text{rms}}^4}{72}$ without affecting the oscillating frequency. An additional very small signal, oscillating at an additional angular frequency $2\omega_s$ also appears for high $\phi_{\text{rms}}$. However, its amplitude is less than $0.5\%$  of the amplitude of the main signal for $\phi_{\text{rms}}\approx1$. The latter has been shown to be optimal for typical Qdyne experiments with statistical polarization \cite{Oviedo2020,Staudenmaier2023prl}. Due to its generality and validity in our specific experimental implementation, we assume the linear approximation further on for simplicity of presentation. 

The expected population of state $|0\rangle_e$ can be detected, e.g. by state-dependent fluorescence by applying a laser pulse. The detected signal is given by \cite{Staudenmaier2022,Staudenmaier2023prl}
\begin{equation}
\label{eq:measurement_signal}
S_k= \eta\left(1+ \frac{c}{2}\langle\phi_0^{(j)}\phi_k^{(j)}\rangle\right)= \eta\left(1+ \frac{c}{2}\phi_{\text{rms}}^2 \cos{\left(\omega_s \Delta T_k \right)}\right)=\eta\left(1+\frac{c}{2}\frac{4}{\pi^2}\gamma_{\text{NV}}^2 B_{\text{rms}}^2t_{\text{DD}}^2\cos{\left(\omega_s \Delta T_k \right)}\right),
\end{equation}
where $\eta=(\eta_0+\eta_1)/2$ is the average signal per measurement with $\eta_0$ ($\eta_1$) the expected signal when the NV electron spin sensor qubit is in state $|0\rangle$ ($|-1\rangle$), while $c=(\eta_0-\eta_1)/\eta$ is the relative experimental contrast. 
We assume that the phase $\phi^{(j)}_k$ is small enough, so that $\sin{(\phi^{(j)}_k)}\approx \phi^{(j)}_k$ for the first equality, and $B_{\text{rms}}^2=B_s^2/2$ for a classical signal with a constant amplitude $g_s=\gamma_{\text{NV}}B_s$ during all measurements (or all NV centers in an ensemble) or $g_{\text{rms}}=\gamma_{\text{NV}}B_{\text{rms}}$, where $B_{\text{rms}}$ is the root mean square of the magnetic field due to statistical polarization \cite{Staudenmaier2022,Staudenmaier2023prl}.
It is evident that the signal obtained after every $k$-th measurement allows us to extract the correlation between the accumulated phase during the $k$-th measurement and during the first initial measurement, which is stored in the nuclear spin populations, with the latter acting as a memory qubit. 
The signal oscillates with an angular frequency $\omega_s$ and we can directly extract the power spectral density of the signal by performing a Fourier transform.
The line width is $\sim 1/T_\text{total}=1/(MT)$, where $M$ is the total number of measurements after the initial phase accumulation and $T=T_{k+1}-T_k$ is the time between two subsequent phase acquisitions. 
The total measurement time $T_\text{total}$ is limited by the lifetime of the nuclear spin memory qubit, considering that it could be affected by the initialization of the electron spin. 

\subsection*{Choice of the Dynamcial Decoupling Sequence}
The simplest dynamical decoupling (DD) sequence is the Hahn echo \cite{Hahn1950}, which consists of a single $\pi$ pulse between the two $\pi/2$ pulses but its $T_2$ is usually not optimal. Robust sequences of multiple $\pi$ pulses like XY8 \cite{Gullion1990jmr}, KDD \cite{RDD_review12Suter}, or the UR family \cite{GenovPRL2017,WangPRL2019} typically perform better for color centers in diamond. These sequences are robust to pulse errors and can tolerate detuning errors of the order of 10 $\%$ of the Rabi frequency (and even more, depending on the sequence and the enrvironmental noise spectrum) with the latter reaching 50-100 MHz with ensembles of NV centers, i.e., inhomogeneous broadening of the order of 5-10 MHz. 

The sequences can also be tailored to NV ensembles by replacing the simple resonant $\pi$ pulses with adiabatic  \cite{GARWOOD1991jmr,Garwood1997adiabatic,GenovPRR2020}, shaped  \cite{Joas2025prx,Daems2013prl,VanDamme2017pra}, composite \cite{Levitt1986,GenovPRL2014} and optimal control \cite{Rembold2020avs,lim2024arxiv} pulses for improved performance. 
For example, robust performance and $T_2$ times close to 2 ms with NV ensembles has been demonstrated experimentally \cite{GenovPRR2020} by combining XY8 with rapid adiabatic passage pulses (RAP-XY8). In this particular experiment, the NV ensemble has a $T_2^*$ of the order of tens of nanoseconds (inhomogeneous broadening of the order of 30 MHz) and amplitude inhomogeneity reaching $40\%$. Typical inhomogeneities are lower than these values with NV ensembles \cite{barry2020sensitivity}, making MCS feasible. We note that MCS benefits from working at high magnetic field, which might lead to a high Larmor frequency of the NMR signal we try to sense. This might require a high duty cycle of the $\pi$ pulses to satisfy the resonance condition, limiting the applicability of pulsed DD. In this case, it is feasible to apply robust continuous DD during the phase acquisition period, e.g., continuous XY8 (CXY8)  \cite{Louzon2025prl}. The latter is robust for amplitude and detuning errors and has been demonstrated to perform well at high magnetic fields.

In this work we use XY8 with rectangular pulses for our proof-of-principle experiment due to its wide applications and simple implementation. We note that the MCS performance can be improved further by using some of the more advanced robust sequences but this goes beyond the scope of this work. 

\section{Comparison of the precision of CS, QDyne, and MCS}
\label{sec:snr_scaling}

In this section, we analyze the precision of the estimation of an unknown frequency $\nu$ of an oscillating classical signal or the Larmor frequency of nuclear spins in a spin bath in the case of nano NMR. 
We define the precision $\delta\nu$ as the uncertainty in estimating the frequency $\nu$. 
In the following, we compare the precision of CS, MCS, and QDyne for a signal due to statistical polarization. 
The analysis is also applicable to sensing the frequency of a classical signal, which phase is not fixed at the beginning of each experiment, i.e., it is not phase locked to the times of the pulses of the applied sensing protocol. 
The precision is determined by the protocols' resolution and signal-to-noise ratio (SNR).

First, we note that, for all three protocols, the resolution is proportional to the inverse of the observed linewidth. In many cases, the resolution of a statistical polarization signal from a liquid sample is determined by the diffusion time \cite{Staudenmaier2022,Staudenmaier2023prl} and is the same for CS, MCS and QDyne. 
This is an intrinsic property of the liquid and can be mitigated by modifying the experimental setup:
Wells can be manufactured inside the diamond that confine the sample liquid \cite{cohen2020confined}.
If the size of the wells is chosen properly, the sample molecules bounce of the walls and reenter the sensing volume, mitigating line broadening by diffusion.
Confinement has already been shown to significantly reduce the diffusion broadening of water signal \cite{zheng2024observation}.
Additionally, fabricating a metal organic framework \cite{liu2022using} on top of the diamond sample can also be utilized for confining the NMR sample, improving the signal's coherence time.

In the following, we analyze the case where diffusion is not the limiting factor. 
Then, the observed line width is determined by $1/T_{\text{total}}$, where $T_{\text{total}}$ is the total duration of the experiment. Both CS and MCS are limited in this aspect by the lifetime of the memory qubit (the nitrogen nuclear spin in our experiment), taking into account the additional decay due to the multiple re-initializations of the sensor qubit (the NV center electron spin in our experiment). The respective limit for MCS is given by 
\begin{equation}
\label{eq:t1_nuc}
\frac{1}{\widetilde{T}_{\text{1,nuc}}}=\frac{1}{T_{\text{1,nuc}}}+\frac{1}{M_\text{limit} T},
\end{equation}
where $T_{\text{1,nuc}}\approx 0.7$ s \cite{Neumann2010singleshot} is the relaxation time of the nitrogen nuclear spin without laser illumination, and $M_\text{limit}=T_{\text{1,nuc,laser}}/t_\text{laser}= \laserrepetitions$ is the number of times we can reinitialize the NV electron spin, so the nuclear spin polarization drops by $1/e$ and $T=T_{k+1}-T_k = \sequencelength$ is the average time between two reinitializations, which depends on the particular chosen sampling time and thus on the undersampling frequency;
$T_{\text{1,nuc,laser}}=\taudecay$ is the relaxation time of the nitrogen nuclear spin during laser illumination with $t_\text{laser} = \taulaser$ the average laser illumination time during one initialization. 
The values are determined by the measurement in \autoref{sec:depolarization}. 
Compared to MCS, CS is typically limited by $T_{\text{1,nuc}}$, as there is no initialization of the electron spin between the two phase acquisition periods, necessary for the repeated readout of MCS.
In contrast to CS and MCS, the line width of QDyne is limited only by the coherence time of the signal, e.g. the diffusion time for liquid nano-NMR, or the coherence time of the clock, which can reach several hundred seconds \cite{SchmittScience2017}. The latter usually makes QDyne the preferred method when sensing a coherent classical signal. 

The other factor determining the precision of a protocol is its SNR. In the following, we assume that photon shot noise is the dominant noise of the protocols, which is typical for experiments with NV centers in diamond. Then, the SNR scales with $T_{\text{total}}^{0.5}$ due to the higher total number of detected photons with a longer total measurement time \cite{SchmittScience2017,Staudenmaier2022,Staudenmaier2023prl}. In the following, we analyze the scaling of the precision and SNR with the total number of NV centers in an ensemble. We note that the precision of frequency estimation is determined by the Cram\`er-Rao bound \cite{Staudenmaier2023prl}:
\begin{equation}\label{eq:precision}
\Delta \omega_s\ge \frac{1}{\sqrt{I(\omega_s)}},
\end{equation}
where $\Delta \omega_s$ is the precision of estimating $\omega_s$ and $I(\omega_s)$ is the Fisher information for estimating the frequency $\omega_s$. 
The latter takes the form \cite{Staudenmaier2023prl}
\begin{align}\label{eq:FI}
I_{\text{CS}}&=\frac{c^2 \eta}{4+c^2 \eta} \phi _{\text{rms}}^4\delta ^2   T_D^3 T_M\propto \frac{c^2 \eta}{4+c^2 \eta}\phi _{\text{rms}}^4 \approx \frac{c^2 \eta}{4}\phi _{\text{rms}}^4 ,\\
I_{\text{QDyne}}&=\left(\frac{c^2 \eta}{4+c^2 \eta}\right)^2\phi _{\text{rms}}^4\frac{T_D^3 T_{\text{total}}\log \left(\omega_u
    T_{\text{total}}\right)}{\Delta T_k^2 }\propto \left(\frac{c^2 \eta}{4+c^2 \eta}\right)^2 \phi _{\text{rms}}^4\approx \left(\frac{c^2 \eta}{4}\right)^2\phi _{\text{rms}}^4,
\end{align}
where $\eta$ is the average photon number per measurement, $c$ is the relative contrast with the last approximation being valid for small relative contrast $c\ll 1$; $\phi_{\text{rms}}=(2/\pi)\gamma_{\text{NV}}B_{\text{rms}}t_{DD}$ is the root-mean-square phase accumulated during the DD sequence, $T_M$ is the lifetime of the memory qubit, which acts as a quantum memory, $T_D$ is the diffusion time in case of liquid nano-NMR, $T_{\text{total}}$ is the total duration of a QDyne experiment, $\Delta T_k=T_k-T_{k-1}$ is the duration of one phase acquisition, $\omega_u=(2\pi)f_u$ is the undersampled angular frequency. 
The Fisher information for MCS, i.e. for the signal in \autoref{eq:measurement_signal}, is similar to that for CS, except for a scaling factor $\propto 2/M$, where $M$ is the number of times sampled, which is related to an improved SNR (see the text below). 
We note that the frequency sensitivity of the protocols is equal to the minimum detectable frequency $\Delta \omega_s$ per unit time, and can be obtained from Eqs.  \eqref{eq:precision} and \eqref{eq:FI}, as described in \cite{DegenRMP2017}. We also note that while the resolution is $\sim 1/T_M$ for CS and MCS, and is inversely proportional to the coherence time of the signal for QDyne, the precision of all protocols is ultimately also affected by the coherence time $T_2$ of the NV sensor as $\phi_{\text{rms}}=(2/\pi)B_{\text{rms}}\tau_{\text{DD}}$ and $\tau_{\text{DD}}<T_2$ \cite{cohen2020confined,Staudenmaier2023prl,DegenRMP2017}. 

We analyze the Fisher information scaling with the number of NV centers $N$ in an ensemble. The expected signal then scales by $\eta\to N\eta$ for all protocols. In the case of CS and MCS, $\phi_{\text{rms}}$ is determined by the average detection volume for each NV center as the signal depends on the correlation $\langle\phi_0^{(j)}\phi_k^{(j)}\rangle$ for each $j$-th NV center, 
i.e., $\phi_{\text{rms}}\to \phi_{\text{rms}}$ for CS and MCS, 
while $\phi_{\text{rms}}\to \phi_{\text{rms}}/\sqrt{N}$ for QDyne with a signal due to statistical polarization, where we detect the average accumulated phase $\langle\phi_k^{(j)}\rangle$ from the detection volume of all NV centers during each measurement, as the individual sensor spins are read out together. 
Thus, the scaling of the Fisher information is given by the ratios 
\begin{align}
R_{\text{CS,MCS}}(N)&=\frac{I_{\text{CS,MCS}}(N)}{I_{\text{CS,MCS}}(1)}\approx  N\,\\
R_{\text{QDyne}}&=\frac{I_{\text{Qdyne}}(N)}{I_{\text{Qdyne}}(1)}=\frac{N^2}{N^2}\approx 1,
\end{align}
It is evident that the Fisher information scales with $\sim N$ for CS and MCS but it stays constant with $N$ for QDyne. As a result, the precision of the experiment with CS and MCS will scale as $1/\sqrt{N}$ and their SNR with $\sqrt{N}$ for ensembles of sensor spins, in contrast to QDyne, where it would remain constant when detecting a signal due to statistical polarization. 
We note that the SNR scaling is different with QDyne when we do the same experiment $N$ times with a single NV center with statistical polarization as $\phi_{\text{rms}}\to \phi_{\text{rms}}$ remains constant during each of the $N$ measurements \cite{SchmittScience2017,Staudenmaier2023prl}. The SNR of QDyne scales as $T_{\text{total}}^{1/2}$ \cite{SchmittScience2017,Louzon2025prl}, which implies $\sim N^{1/2}$ scaling of SNR when $N$ is the number of measurements. Similarly, when we detect a classical signal or signal due to thermal polarization with an NV ensemble $\phi_{\text{rms}}$ stays constant with $N$, where $N$ is the number of NV centers in the ensemble,  and the SNR scaling of QDyne and other related protocols is also $\sim N^{1/2}$ \cite{yoon2025quantum,Glenn2018CASR}.

Finally, the SNR of MCS is improved in comparison to that of CS due to the reduction of measurement idle time.  
In MCS, $M$ phase measurements of phases $\phi_k$ (here, we drop the subscript $(j)$ for simplicity of presentation), which are correlated with an initial phase $\phi_0$ are acquired subsequently within a single sequence run.
The total measurement time to acquire one set of data is thus given by
\begin{equation}
    T_\mathrm{tot, MCS} = T_\mathrm{init} + M T\,,
\end{equation}
where $T_\mathrm{init}$ is the time needed for initializing the system and acquiring the initial phase $\phi_0$ (steps I and II in Figure 1 of the main text), which is the same for CS and MCS, and $T = T_{k+1} - T_k$ is the time between two subsequent phase measurements $\phi_k$ and $\phi_{k+1}$ (step III in Figure 1 of the main text).

In CS, each measurement only produces the resulting correlation of one phase $\phi_k$ to an initial phase $\phi_0$.
To acquire the complete autocorrelation function of all $M$ phases, separate measurements must be performed for each phase $\phi_k$.
During these measurements, the time between the initial and second phase acquisitions $\Delta T_k = T_k - T_0 = k T$ is swept.
This results in a total measurement time
\begin{align}
    T_\mathrm{tot, CS} &= \sum_{k=1}^M \left(T_{\text{init}}+k T\right) = M T_{\text{init}} + \frac{M (M + 1)}{2}  T\,.
\end{align}
 The ratio between the total measurement times $T_\mathrm{tot, CS}$ and $T_\mathrm{tot, MCS}$ is thus
\begin{align*}
    \frac{T_\mathrm{tot, CS}}{T_\mathrm{tot, MCS}} &= M \frac{1 + \frac{M+1}{2} \frac{ T}{T_{\text{init}}}}{1 + M \frac{ T}{T_\text{init}}} = \frac{M}{2}\left(1+\frac{1+T/T_{\text{init}}}{1+M T/T_{\text{init}}}\right) \approx \measurementtimefactorcs\,,
\end{align*}
where the final equality is given for the parameters of $ T = \sequencelength$, $M=\phaseacqreps$ and $T_\text{init} = \Tzero$, used in the experiments. Assuming that photon shot noise is the dominant noise in the experiment, the SNR of MCS will improve in comparison to CS with a factor
\begin{equation}
\label{eq:f_T}
f_T = \sqrt{\frac{T_\mathrm{tot, CS}}{T_\mathrm{tot, MCS}}}= \sqrt{\frac{M}{2}\left(1+\frac{1+T/T_{\text{init}}}{1+M T/T_{\text{init}}}\right)} \approx \sqrt{\measurementtimefactorcs}\approx 31.6.
\end{equation}

\section{Nitrogen polarization during laser illumination}
\label{sec:depolarization}

\begin{figure}
    \centering
    \includegraphics[scale=1]{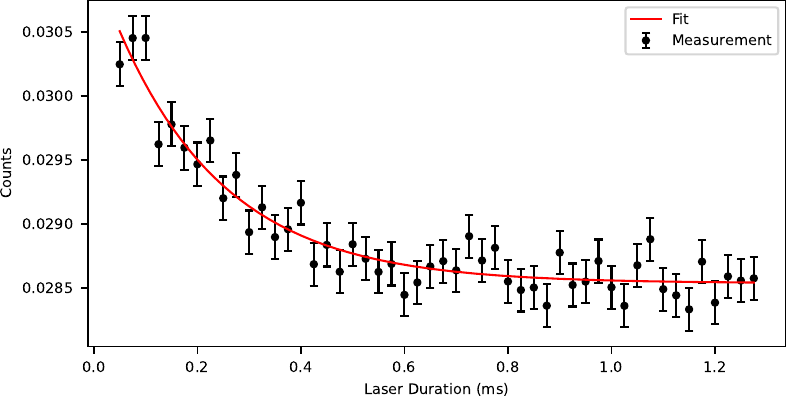}
    \caption{Nitrogen nuclear spin polarization over time during repeated laser illumination.
    The nuclear spin is initialized using the initialization step in \autoref{SM_Subsection:Initialization} and its polarization repetitively read out by applying a selective MW pulse and laser readout.
    A decay time $T_{1, \mathrm{nuc, laser}} = \taudecay$ for the nuclear spin under laser illumination was detected.
    The measurement was acquired for $\sim \SI{57}{\min}$ with $1017037$ measurement sweep repetitions.}    \label{fig:nuclear_init_decay}
\end{figure}

In order to get an estimate of how many readout lasers $M$ can be performed before the nuclear spin population has decayed and the correlation to the initially acquired phase $\phi^{(j)}_0$ can no longer be performed, the readout laser is repetitively applied.
Before each readout laser, the nuclear spin state is retrieved by applying a $\cnotn$ gate.
The resulting fluorescence response is plotted in \autoref{fig:nuclear_init_decay} against the total laser duration $M t_\mathrm{laser}$, where $t_\mathrm{laser}$ is the readout laser length.
The data is fit by using
\begin{equation}
    f(x) = b + a \exp \left( - \frac{x}{T_{\text{1,nuc,laser}}} \right)\,,
\end{equation}
where $a$ is the amplitude of the exponential at $x=0$, $T_{\text{1,nuc,laser}}$ is the nitrogen polarization decay time during laser illumination and $b$ a general offset of the data.
The fit yields $T_{\text{1,nuc,laser}} = \taudecay$, which corresponds to $M_\mathrm{limit} = \laserrepetitions$ repetitions with a laser length $t_\mathrm{laser} = \taulaser$.
Therefore, the laser used to reinitialize the NV center during the initialization step in \autoref{SM_Subsection:Initialization} mainly leaves the built-up nuclear polarization untouched and a sufficient number of phase acquisitions can be performed.

%% file: tex/supp/supp_measurement.tex
\section{Experimental Setup \& Parameters}

Measurements were performed on a home-built confocal microscope, using a natural abundance $^{13}$C bulk diamond sample.

Through an objective (Olympus PlanApo N 60x Oil Microscope), the laser is focused onto the NV center, which is positioned using a nanopositioner (PI P-562.3CD).
Its fluorescence response is captured by the same objective and the emitted photons are measured using an avalanche photo diode (Excelitas SPCM-AQRH-4X-TR) and counted using a counting card (FastComtec MCS6A).
The control of electron and nuclear spin is performed by spanning a copper wire with a diameter of $\SI{20}{\micro \meter}$ close to the NV center.
Through the same wire, a test signal is supplied that is generated by a signal generator (Rohde \& Schwarz SMIQ06B).
The pulse sequence is orchestrated with the help of an arbitrary waveform generator (AWG) (Keysight M8915A).
The MW and RF signals are amplified (AR 50W1000 and AR 30S1G6, respectively) before reaching the sample.
The test signal and RF control are combined using a power splitter (Minicircuits ZFRSC-2050+) and the different frequency bands are combined using a diplexer (Amphenol Procom PRO-DIPX 520/790-2700).
For NV center electron spin initialization, a $\SI{515}{\nano \meter}$ (Toptica iBeam-Smart-515-SA3-14710) is used, which can be pulsed with a TTL signal from the AWG.
The experiment was controlled using the Qudi \cite{Binder2017qudi} software suite.
The static bias magnetic field $B_0$ is supplied by a home-built permanent magnet and aligned with the NV center symmetry axis.
Alignment was performed using the method in \cite{Liu2019}.
The magnetic field strength of $\bzero$ was determined from the center frequency of a pulsed ODMR spectrum, as shown in \autoref{fig:nuclear_init}.

\begin{figure}
    \centering
    \includegraphics[scale=1]{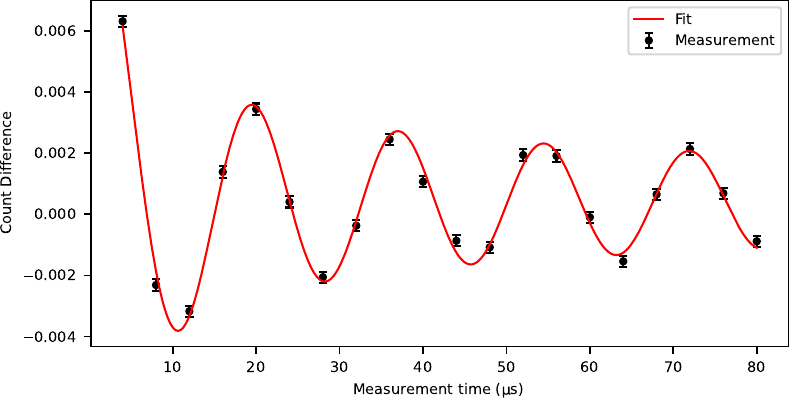}
    \caption{
    XY8 order sweep with fixed inter-pulse spacing of $\tau = \xytau$, while a $\signalfrequency$ RF signal is applied.
    The resulting NV fluorescence response is fit using \autoref{eq:bessel}.
    A signal strength of $\testsignalstrength$ at the NV center location was found.
    The measurement was acquired for $\sim \SI{1}{\hour}$ with $1956310$ measurement sweep repetitions.}
    \label{fig:xy8}
\end{figure}

The strength of the $\signalfrequency$  RF signal at the location of the NV center was determined by performing an XY8 measurement with the waiting time between pulses set to the resonance of the signal with $\tau=\xytau$.
The number of repetitions $N$ of the XY8 block is swept and the resulting fluorescence response is plotted against the total measurement time $t_\mathrm{meas} = 8 N \tau$ in \autoref{fig:xy8}.
For a signal without phase lock, the data should follow a Bessel function of the first kind $J_0$ \cite{DegenRMP2017}.
Therefore, the data is fit with 
\begin{equation}
    f(x) = a J_0(2 \pi f x)\,,
    \label{eq:bessel}
\end{equation}
where $a$ is the amplitude and $f$ the frequency of the oscillation.
The fit yields a signal strength of $B_\mathrm{test} = \testsignalstrength$ at the NV center position.

\subsection{Measurement Information}

All presented measurements were acquired by counting the photons detected by the APD.
These photon counts are then summed up for each measurement point for the duration of the experiment.
The graphs in the main article and in the supplementary material show the number of photons acquired per measurement (laser) and are labeled as counts.
This value is calculated by dividing the total number of acquired photons per measurement point by the number of measurement sweeps.
The data is presented in this way to mitigate the effect of different measurement accumulation periods.
The difference in counts per measurement in the different figures is originating from varying laser power between the measurements as laser power was not explicitly kept constant between all of them.

In \autoref{fig:xy8} and in \autoref{fig:nuclear_odmr} the count difference is plotted.
This is the difference in number of photons acquired per readout when subtracting the measurement from the alternating measurement.
The alternating measurement is recorded by appending an MW $\pi$-pulse before each readout.
This is done to mitigate errors caused by fluctuations in e.g. the charge state and count rate \cite{Staudenmaier2022, Louzon2025prl}.
In \autoref{fig:nuclear_odmr} this is done to avoid waste of measurement time as each data point has to be followed by a depolarization step, reverting the polarization step, to not build up polarization over the measurement sweep.

\section{Multipoint Correlation Spectroscopy Data Analysis}

The time trace of the acquired data in Figure 3 (a) of the main article is fit by
\begin{equation}
    f(x) = b + a \sin\left( 2 \pi f x + \varphi \right) \exp \left( - \frac{x}{\tau_\mathrm{decay}} \right) \,,
\end{equation}
where $a$ is the amplitude, $f$ the frequency and $\varphi$ the phase of the oscillation.
Furthermore, an offset $b$ and an exponential decay with the decay time $\tau_\mathrm{decay}$ of the signal were considered. 

As the obtained signal is directly related to the autocorrelation function of the accumulated phases during two measurements \cite{DegenRMP2017,Staudenmaier2022}, it is directly related to the autocorrelation function of the magnetic fields at different times. Then, taking the FFT of the CS/MCS signal directly corresponds to taking the FFT of the autocorrelation function of the magnetic field to be sensed. As a result, by taking the FFT of the CS/MCS signal we obtain the power spectral density (PSD) of the sensed oscillating magnetic field. 

The power spectral density of the signal in Figure 2 (b) of the main text is fit by the Lorentzian function
\begin{equation}\label{eq:fft}
    f(x) = b + a \frac{\sigma^2}{\left( x - c \right)^2 + \sigma^2}\,,
\end{equation}
where $a$ is the amplitude, $2\sigma$ the full width at half maximum, $c$ the center frequency of the peak and $b$ a general offset.

Sensitivity was determined by comparing the signal strength with the noise level in the power spectral density of the measurement data of Figure 2 (b) of the main article.
The noise level $\sigma_\mathrm{noise}$ was recovered by subtracting the fit result from the data and calculating the standard deviation.
The sensitivity is then calculated by
\begin{equation}
    \eta_\mathrm{sens} = \frac{\sigma_\mathrm{noise}}{f(c)} B_\mathrm{test} \sqrt{T_\mathrm{meas}} \,,
\end{equation}
where $f(c)$ is the fit result of \autoref{eq:fft} at the signals frequency, $B_\mathrm{test}$ is the magnetic field of the test signal at the NV center position and $T_\mathrm{meas}$ is the total measurement time.
The measurement was acquired for $\measurementtime$, which results in a sensitivity of $\sensitivity$.
This value could be optimized by repeating the phase acquisition block fewer times.
Furthermore, optimizations in the NV center, and thus nuclear spin initialization, could further boost the sensitivity.

\section{Influence of the Bias Magnetic Field}
\begin{figure}
    \centering
    \includegraphics[scale=1]{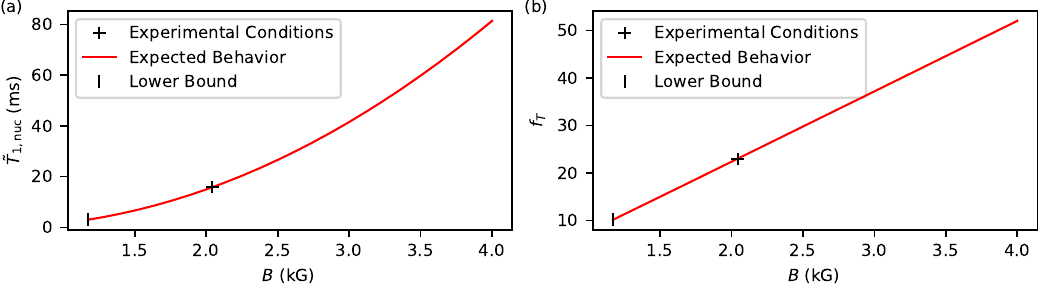}
    \caption{(a) Nitrogen nuclear lifetime $\widetilde{T}_{1, \mathrm{nuc}}$ for different magnetic field strengths $B$ calculated using \autoref{eq:quadratic_b}.
    Furthermore, the experimental conditions as used in the measurements are marked as well as the lower bound, where the nuclear lifetime surpasses $\SI{3}{\milli \second}$.
    (b) The resulting SNR improvement $f_T$ of the MCS protocol over CS, calculated from \autoref{eq:f_T}.
    The number of phase acquisitions was chosen as $M(B) = \widetilde{T}_{1, \mathrm{nuc}}(B) / T$, where $T$ is the length of one phase acquisition block and $\widetilde{T}_{1, \mathrm{nuc}}(B)$ is taken from the data in (a).
    The lower bound already shows a SNR improvement of MCS over CS of $\sim 10 \times$.
    }
    \label{fig:nitrogen_t1_ft}
\end{figure}

An increase in the magnetic field strength $B$ leads to a quadratic increase in the nitrogen nuclear lifetime $\widetilde{T}_{1, \mathrm{nuc}}$ \cite{Neumann2010singleshot}.
\autoref{fig:nitrogen_t1_ft} (a) shows the nuclear spin lifetime $\widetilde{T}_{1, \mathrm{nuc}}$ for our sample and experimental conditions during the MCS measurement.
Using this known value, we can determine the behavior of $\widetilde{T}_{1, \mathrm{nuc}}$, following \cite{Neumann2010singleshot}, to be
\begin{equation}
    \label{eq:quadratic_b}
    \widetilde{T}_{1, \mathrm{nuc}} (B) = \SI{0.6645}{\s \per \T \squared} \left( B - \SI{50}{\milli \tesla} \right) ^2
\end{equation}
and thus determine the expected $\widetilde{T}_{1, \mathrm{nuc}}$ for different magnetic fields $B$.
This nuclear spin lifetime is specific to the sample used and the exact parameters of the MCS measurement.
For instance, by decreasing the number of phase acquisitions $M$, fewer laser readouts are performed, and thus a longer $\widetilde{T}_{1, \mathrm{nuc}}$ can be expected (see \autoref{eq:t1_nuc}).

The improvement in SNR of the MCS protocol over CS ($f_T$ from \autoref{eq:f_T}), due to the temporal advantage, is depicted in \autoref{fig:nitrogen_t1_ft} (b). 
The number of phase acquisitions is chosen as $M(B) = \widetilde{T}_{1, \mathrm{nuc}} (B) / T$, where $T=\SI{15.06}{\micro \second}$ is the length of one phase acquisition block.
The improvement in SNR shows a linear dependence on the magnetic field strength.
Therefore, for coherent signals the highest possible magnetic field yields the highest SNR improvement of MCS over CS.

A lower bound for the required magnetic field can be estimated when considering common NV center electron spin $T_1$ times.
In order to achieve a higher frequency resolution with the MCS protocol or memory assisted CS over standard CS the nuclear spin should have a longer $\tilde{T}_{1, \mathrm{nuc}}$ time than the electron spin of the NV center.
Common NV electron spin $T_1$ times are $\sim \SI{3}{\milli \second}$ \cite{jarmola2015longitudinal}.
At magnetic fields of more than $\sim \SI{1172}{\gauss}$ the nuclear spin $\tilde{T}_{1, \mathrm{nuc}}$ exceeds that of the electron spin under our experimental conditions. At this lower bound, MCS is already preferable to using standard correlation spectroscopy without utilizing a memory qubit by a factor of $10 \times$.

The application of a strong magnetic field might lead to a lower contrast with an ensemble of NV centers because of inhomogeneities, diminishing the effect of an SNR increase.
Generally however, NV ensemble measurements can offer a contrast exceeding $\SI{8}{\percent}$ \cite{vetter2024gate} and single centers can offer a contrast of $\sim \SI{30}{\percent}$ \cite{vetter2022zero}   .
The SNR of sensing experiments (and our protocol) changes $\propto C$ with $C$ being the contrast of the measurement as described in \autoref{sec:snr_scaling}.
This means that due to the loss of contrast in an ensemble, the SNR is reduced by roughly a factor of $3$ compared to the single NV, when everything else is kept constant. 
However, the SNR increases $\propto \sqrt{N}$ with $N$ being the number of NV's as described in \autoref{sec:snr_scaling}.
For example, assuming that we have $50$ NV centers in the confocal volume, the SNR increases by a factor of approximately $7$ compared to a single NV. 
Combining the loss in SNR due to contrast loss with the increase in SNR due to NV number increase, the SNR improves by the factor $\frac{7}{3} \approx 2.3$.
By using more NV centers, e.g., by using a larger detection volume or a more dense ensemble, this improvement can be enhanced even further.

\section{Limits of Frequency Resolution for Signals due to Statistical Polarization}

\begin{figure}
     \centering
     \includegraphics[width=\linewidth]{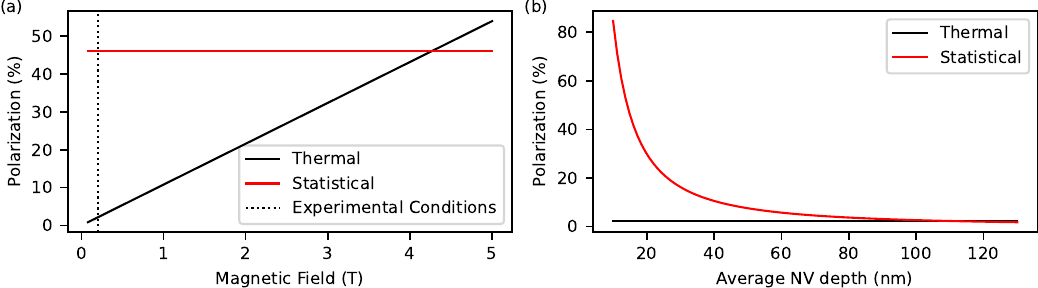}
     \caption{\textsuperscript{1}H  nuclear spin polarization for thermal and statistical polarization calculated following \textcite{herzog2014boundary}. 
     (a) Polarization dependence on the magnetic field.
     The depth for the statistical polarization was assumed to be $\SI{15}{\nano \meter}$.
     Thermal polarization surpasses statistical polarization above $\sim \SI{4}{\tesla}$.
     (b) Polarization dependence on the average NV center depth.
     The magnetic field for the thermal polarization corresponds to the magnetic field $B_z = \bzero$, used in the manuscript.
     Thermal polarization surpasses statistical polarization for average NV depths of more than $\sim \SI{100}{\nano \meter}$.
     }
     \label{fig:polarization}
 \end{figure}

Apart from improving the signal-to-noise ratio, frequency resolution is also an important factor 
in nuclear magnetic resonance experiments. We note that alternative methods that rely on thermal polarization, such as CASR \cite{Glenn2018CASR} or QDyne \cite{SchmittScience2017}, are not limited by the lifetime of the memory qubits, i.e., the nuclear $T_{1, \mathrm{nuc}}$ limit. Thus, they can in principle provide improved frequency resolution in comparison to MCS if the signal-to-noise ratio is high enough and the coherence time is not limited by other factors, e.g., magnetic field stability. However, our protocol targets a complementary parameter regime in which statistical polarization is the dominating signal contribution.
CASR/QDyne are preferably applicable for signals due to thermal polarization, while MCS is preferable for detecting a signal due to statistical polarization, although it can be used for both types of signals.  
Specifically, in our measurement setting and at the particular applied magnetic field the statistical \textsuperscript{1}H  nuclear spin polarization is expected to be significantly higher than the thermal polarization, as shown in \autoref{fig:polarization} (a). 
In the figure, we assume an average NV center depth of $\SI{15}{\nano \meter}$, which is common for NV centers implanted with $\SI{5}{\kilo e \volt}$ \cite{fuhrmann2025photoactivation}, commonly used for nanoscale NMR experiments \cite{Staudenmaier2023prl}.
 The nuclear spin polarization of the sample liquid has been calculated following \textcite{herzog2014boundary}.
 If the signal-to-noise ratio is high enough and the signal can be detected with such weak polarization, then CASR/QDyne may yield an improvement in frequency resolution. 
 Of course, if larger nuclear spin polarization is required, one can also choose a higher magnetic field, which increases the nuclear spin lifetime and thus improves frequency resolution \cite{maier2025}. However, utilizing higher magnetic fields is sometimes not possible or undesirable, leaving MCS detection of statistically polarized signal as the only available technique in case of insufficient signal-to-noise ratio with thermal polarization. Overall, MCS and CASR/QDyne are designed to solve different problems and to be complementary when used with NV ensembles. 

The frequency resolution of signals due to statistical polarization is also limited by diffusion \cite{Staudenmaier2022,Staudenmaier2023prl}. The effects of diffusion broadening can, in principle, be mitigated by increasing the depth of the NV centers, as this increases the effective detection volume, as stated in the main manuscript.
However, as shown in \autoref{fig:polarization} (b), the statistical polarization of the \textsuperscript{1}H nuclear spin sample decreases significantly with a larger average depth of the NV centers.
Nevertheless, even if we double the average depth of the NV centers to, e.g., 30 nm, there would be more than $\sim \SI{10}{\percent}$ statistical polarization, which should be $\sim 5 \times$ larger than the thermal polarization with the measurement settings of the main paper. Thus, depending on signal strength and coherence, as well as readout efficiency of the sensing spins in a given setting, careful optimization of the NV depth is required to achieve the best possible frequency resolution.
Then, the effect of diffusion broadening can be reduced, allowing for improved frequency resolution at the expense of lower signal-to-noise ratio. 

Nevertheless, we note that alternative methods for reducing diffusion broadening, e.g. confinement \cite{cohen2020confined, liu2022using, zheng2024observation}, despite its experimental complexity, might be a more preferable approach towards limiting diffusion broadening, as the signal strength is likely to remain unaffected.